\documentclass[prd,reprint,longbibliography,nofootinbib,superscriptaddress]{revtex4-2}

\usepackage[USenglish]{babel}
\usepackage[T1]{fontenc}
\usepackage{amsmath}
\usepackage{mathtools}
\usepackage{amssymb}
\usepackage{graphicx}
\usepackage{physics}
\usepackage[acronym]{glossaries}
\usepackage{siunitx}
\usepackage[svgnames]{xcolor}
\usepackage[hidelinks]{hyperref}
\hypersetup{
  colorlinks = true, 
  urlcolor = DarkBlue, 
  linkcolor = DarkBlue, 
  citecolor = DarkBlue  
}
\usepackage{cleveref}
\creflabelformat{equation}{#2\textup{#1}#3}

\DeclareMathOperator*{\argmin}{arg\,min}

\begin{document}

\title{Ranging Sensor Fusion in LISA Data Processing: Treatment of Ambiguities, Noise, and On-Board Delays in LISA Ranging Observables}

\author{Jan Niklas Reinhardt}
\email{janniklas.reinhardt@aei.mpg.de}
\affiliation{Max-Planck-Institut für Gravitationsphysik (Albert-Einstein-Institut),\\ Callinstraße 38, 30167 Hannover, Germany}
\affiliation{Leibniz Universität Hannover, Welfengarten 1, 30167 Hannover, Germany}

\author{Martin Staab}
\affiliation{Max-Planck-Institut für Gravitationsphysik (Albert-Einstein-Institut),\\ Callinstraße 38, 30167 Hannover, Germany}
\affiliation{Leibniz Universität Hannover, Welfengarten 1, 30167 Hannover, Germany}

\author{Kohei Yamamoto}
\affiliation{Max-Planck-Institut für Gravitationsphysik (Albert-Einstein-Institut),\\ Callinstraße 38, 30167 Hannover, Germany}
\affiliation{Leibniz Universität Hannover, Welfengarten 1, 30167 Hannover, Germany}

\author{Jean-Baptiste Bayle}
\affiliation{University of Glasgow, Glasgow G12 8QQ, United Kingdom}

\author{Aurélien Hees}
\affiliation{SYRTE, Observatoire de Paris, Université PSL, CNRS, Sorbonne Université,\\LNE, 61 avenue de l'observatoire 75014 Paris, France}

\author{Olaf Hartwig}
\affiliation{Max-Planck-Institut für Gravitationsphysik (Albert-Einstein-Institut),\\ Callinstraße 38, 30167 Hannover, Germany}
\affiliation{Leibniz Universität Hannover, Welfengarten 1, 30167 Hannover, Germany}
\affiliation{SYRTE, Observatoire de Paris, Université PSL, CNRS, Sorbonne Université,\\LNE, 61 avenue de l'observatoire 75014 Paris, France}

\author{Karsten Wiesner}
\affiliation{Max-Planck-Institut für Gravitationsphysik (Albert-Einstein-Institut),\\ Callinstraße 38, 30167 Hannover, Germany}
\affiliation{Leibniz Universität Hannover, Welfengarten 1, 30167 Hannover, Germany}

\author{Sweta Shah}
\affiliation{Max-Planck-Institut für Gravitationsphysik (Albert-Einstein-Institut),\\ Callinstraße 38, 30167 Hannover, Germany}
\affiliation{Leibniz Universität Hannover, Welfengarten 1, 30167 Hannover, Germany}

\author{Gerhard Heinzel}
\affiliation{Max-Planck-Institut für Gravitationsphysik (Albert-Einstein-Institut),\\ Callinstraße 38, 30167 Hannover, Germany}
\affiliation{Leibniz Universität Hannover, Welfengarten 1, 30167 Hannover, Germany}


\begin{abstract}
Interspacecraft ranging is crucial for the suppression of laser frequency noise via time-delay interferometry (TDI). So far, the effects of on-board delays and ambiguities on the LISA ranging observables were neglected in LISA modelling and data processing investigations. In reality, on-board delays cause offsets and timestamping delays in the LISA measurements, and pseudo-random noise (PRN) ranging is ambiguous, as it only determines the range up to an integer multiple of the PRN code length. In this article, we identify the four LISA ranging observables: PRN ranging, the sideband beatnotes at the interspacecraft interferometer, TDI ranging, and ground-based observations. We derive their observation equations in the presence of on-board delays, noise, and ambiguities. We then propose a three-stage ranging sensor fusion to combine these observables in order to gain accurate and precise ranging estimates. We propose to calibrate the on-board delays on ground and to compensate the associated offsets and timestamping delays in an initial data treatment (stage 1). We identify the ranging-related routines, which need to run continuously during operation (stage 2), and implement them numerically. Essentially, this involves the reduction of ranging noise, for which we develop a Kalman filter combining the PRN ranging and the sideband beatnotes. We further implement crosschecks for the PRN ranging ambiguities and offsets (stage 3). We show that both ground-based observations and TDI ranging can be used to resolve the PRN ranging ambiguities. Moreover, we apply TDI ranging to estimate the PRN ranging offsets.
\end{abstract}

\maketitle

\section{Introduction}\label{sec:intro}
The Laser Interferometer Space Antenna (LISA), due for launch around the year 2035, is an ESA-led mission for space-based gravitational-wave detection in the frequency band between \SI{0.1}{\milli \hertz} and \SI{1}{\hertz} \cite{Amaro2017LISA}. LISA consists of three satellites forming an approximate equilate\-ral triangle with an armlength of \SI{2.5}{\giga\m}, in a heliocentric orbit that trails or leads Earth by about \num{20} degrees. Six infrared laser links with a nominal wavelength of \SI{1064}{\nano\m} connect the three spacecraft (SC), whose relative motion necessitates the usage of heterodyne interferometry. Phasemeters are used to extract the phases of the corresponding beatnotes \cite{Gerberding:Phasemeter}, in which gravitational-waves manifest in form of microcycle deviations equivalent to picometer variations in the interspacecraft ranges.\par
\begin{figure}
	\begin{center}
		\includegraphics[width=0.48\textwidth]{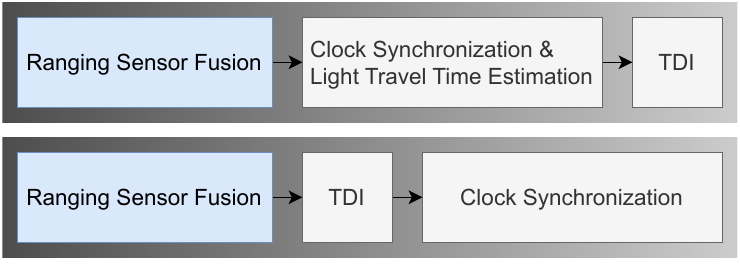}
	\end{center}
	\caption{In the baseline TDI topology (upper part) we perform TDI after clock synchronization to TCB, the delays are given by the light travel times. In the alternative TDI topo\-logy (lower part) we execute TDI without clock synchronization and apply the pseudoranges as delays \cite{Hartwig:TDIwoSync}. Both topologies rely on a ranging sensor fusion.}\label{fig:tdi-topologies}
\end{figure}

The phasemeter output, however, is obscured by various instrumental noise sources. They must be suppressed to fit in the LISA noise budget of \SI{10}{\pico \m \hertz \tothe{-0.5}} (single link) \cite{barkeLISAMetrologySystem}, otherwise they would bury the gravitational-wave signals. Dedicated data processing algorithms are being developed for each of these instrumental noise sources, their subsequent execution is referred to as initial noise reduction pipeline (INReP). The dominating noise source in LISA is by far the laser frequency noise, which must be reduced by more than 8 orders of magnitude. This is achieved by time-delay interferometry (TDI), which combines the various beatnotes with the correct delays to virtually form equal-optical-path-length interferometers, in which laser frequency noise naturally cancels \cite{ArmstrongTDI, TintoTDIforLISA}. The exact definition of these delays depends on the location of TDI within the INReP (see \cref{fig:tdi-topologies}) \cite{Hartwig:TDIwoSync}, but wherever we place it, some kind of information about the absolute interspacecraft ranges is required.\par
Yet, absolute ranges are not a natural signal in a continuous-wave heterodyne laser interferometer such as LISA. Therefore, a ranging scheme based on pseudo-random noise (PRN) codes is implemented \cite{EstebanPRN1,EstebanPRN2, Heinzel-Ranging}: each SC houses a free-running ultra-stable oscillator (USO) as timing reference. It defines the spacecraft elapsed time (SCET). PRN codes generated according to the respective SCETs are imprinted onto the laser beams by phase-modulating the carrier. The comparison of a PRN code received from a distant SC, hence generated according to the distant SCET, with a local copy enables a measurement of the pseudorange: the pseudorange is commonly defined as the difference between the SCET of the recei\-ving SC at the event of reception and the SCET of the emitting SC at the event of emission \cite{hartwig2021instrumental}. It represents a combination of the true geometrical range (light travel time) with the offset between the two involved SCETs (see \cref{eq-ap:Pseudorange-TCB}).\par
In the baseline TDI topology (upper row in \cref{fig:tdi-topologies}), TDI is performed after SCET synchronization to the barycentric coordinate time (TCB), the light travel times are used as delays. The pseudoranges comprise information about both the light travel times and the SCET offsets required for synchronizing the clocks (see \cref{appendix:pr_TCB}). A Kalman filter can be used to disentangle the pseudoranges in order to retrieve light travel times and SCET offsets \cite{Wang:KF}. In the alternative TDI topology (lower row in \cref{fig:tdi-topologies}), the pseudoranges are directly used as delays. In that topology, TDI is executed on the unsynchronized beatnotes sampled according to the respective SCETs \cite{Hartwig:TDIwoSync}.\par
However, PRN ranging (PRNR) does not directly provide the pseudoranges but requires three treatments. First, due to the finite PRN code length (we assume \SI{400}{\kilo\m}), PRNR measures the pseudoranges modulo an ambiguity \cite{EstebanPRN1}. Secondly, on-board delays due to signal propagation and processing cause offsets and time\-stamping delays in the PRNR. Thirdly, PRNR is li\-mited by white ran\-ging noise with an rms amplitude of about \SI{0.3}{\m}, which is due to shot noise and PRN code interference \cite{Heinzel-Ranging, sutton2010laser}. To overcome these difficulties, there are three additional pseudo\-range observables, which are actually designed for other purposes and serve that function secondarily: ground-based observations provide inaccurate but unambiguous pseudorange estimates; time-delay interfero\-metric ranging (TDIR) turns TDI upside-down seeking a model for the delays that minimizes the laser frequency noise in the TDI combinations \cite{TintoTDIR}; the sideband beatnotes, primarily designed for clock noise correction, provide a measurement of the pseudorange time derivatives \cite{Hartwig:TDIwoSync}. The combination of these four observa\-bles in order to form optimal pseudorange estimates is referred to as \textit{ranging sensor fusion} in the course of this article. It is common to both TDI topologies (see \cref{fig:tdi-topologies}) and consequently a crucial stage of the INReP.\par
In \cref{sec:pseudorange-and-on-board-delays}, we specify the pseudorange definition in the context of on-board delays and identify the delays required for TDI. We then derive observation equations for the four pseudorange observables in \cref{sec:ranging-measurements}. Here, we carefully consider the effects of on-board delays at the interspacecraft interferometer.\footnote{We neglect such delays for the reference and test-mass interfe\-rometers, which we will treat in a follow-up work.} In \cref{sec:sensor-fusion}, we introduce a three-stage ranging sensor fusion consisting of an initial data treatment, a core ranging processing, and crosschecks. In the initial data treatment, we propose to compensate for the offsets and timestamping delays caused by the on-board delays. We identify PRNR unwrapping and noise reduction as the ranging processing steps that need to run continuously during operation. In parallel to this core ranging processing, we propose crosschecks of the PRNR ambiguities and offsets. We implement the core ranging processing and the crosschecks numerically. In \cref{sec:results} we discuss the performance of this implementation, and conclude in \cref{sec:conclusion}. 

\section{The pseudorange and TDI in the context of on-board delays}\label{sec:pseudorange-and-on-board-delays}
\subsection{Brief description of the LISA payload}
\begin{figure}
	\centering
	\includegraphics[width=0.45\textwidth]{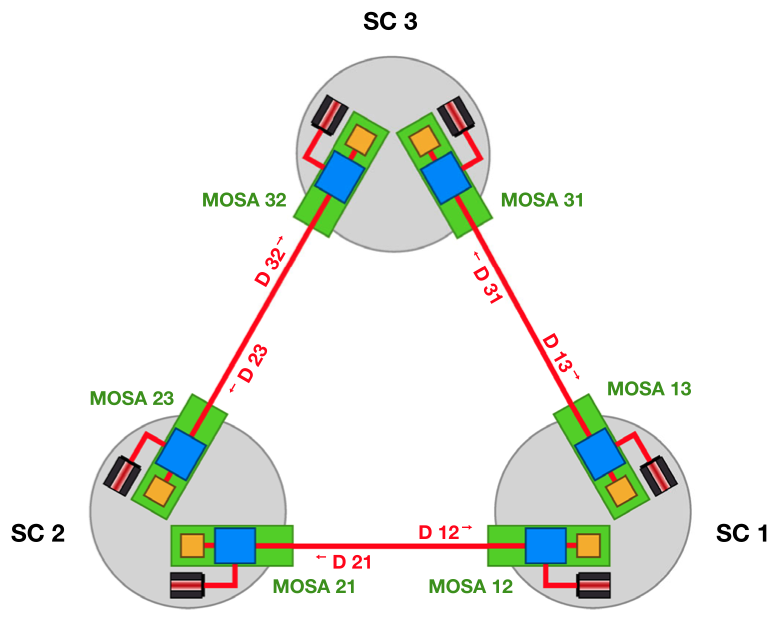}
	\caption{LISA labeling conventions (from \cite{Bayle:TDI-in-Frequency}). The SC are labeled clockwise. The MOSAs and their associated building blocks (lasers, interferometers, etc.) are labeled by 2 indices: the first one indicates the SC they are located at, the second one the SC they are oriented to. The measurements and related quantities (optical links, pseudoranges, etc.) share the indices of the MOSAs they are measured at. Below, we distinguish between left-handed MOSAs (12, 23, 31) and right-handed ones (13, 32, 21).}
	\label{fig:LISA-Notation}
\end{figure}
Each SC houses an ultra-stable oscillator (USO) gene\-rating a signal at about \SI{50}{\mega\hertz}. An \SI{80}{\mega\hertz} clock signal, the phasemeter clock (PMC), is coherently derived from this USO. The PMC can be considered as the timing refe\-rence on board the SC (see \cref{fig:LISA-metrology-system}), its associated counter is referred to as spacecraft elapsed time (SCET):
\begin{align}
	\text{SCET}(n)=\sum_{1}^{n}\: \frac{1}{\SI{80}{\mega\hertz}}.
\end{align}
It is useful to consider the SCET as a continuous time scale, which we denote by $\hat{\tau}$. It differs from the barycentric coordinate time (TCB), denoted by $t$, due to instrumental clock drifts and jitters, and due to relativistic effects. Following the notation of \cite{Bayle:LISA-Simulation}, we use superscripts to indicate a quantity to be expressed as function of a certain time scale, e.g., $\hat{\tau}_1^{t}$ denotes the SCET of SC 1 as function of TCB. Note that
\begin{align}\label{eq:SCET-in-SCET}
	\hat{\tau}_{i}^{\hat{\tau}_{i}}(\tau) = \tau.
\end{align}
\begin{figure*}
	\begin{center}
		\includegraphics[width=1\textwidth]{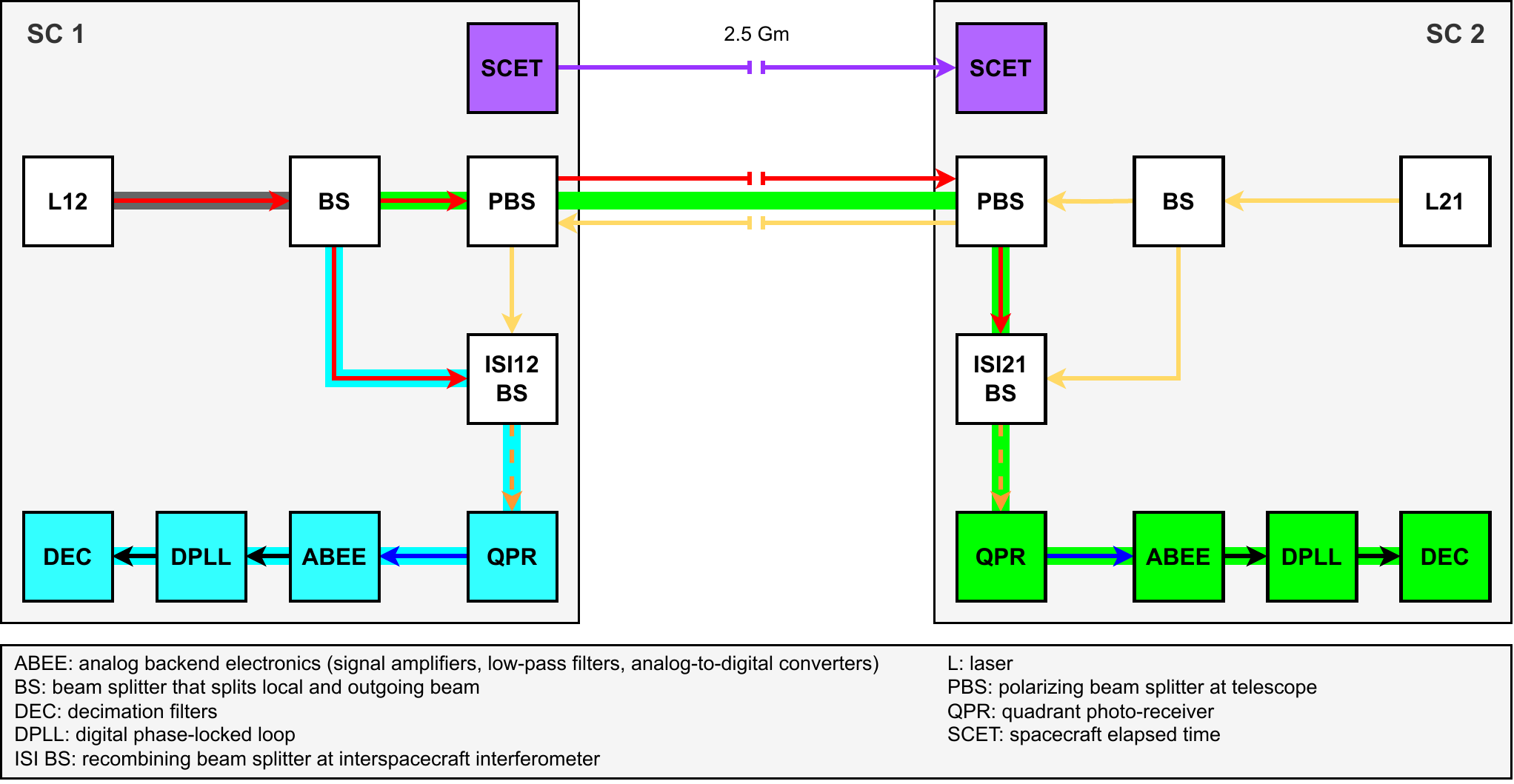}
	\end{center}
	\caption{We trace laser 12 (red arrows) and laser 21 (yellow arrows) to the ISI BSs on both SC, where they interfere and form beatnotes (orange dashed arrows). In the subsequent data processing chain we indicate analog signals by blue and digital signals by black arrows. Constituents of the pseudorange are marked purple. These are the light travel time between the PBSs (at the telescopes) and the transformation between the two SCETs. In \cref{sec:delays-for-TDI} we identify the TDI delay that is required to cancel the noise of laser 12. For that purpose we mark the common path of the outgoing and the local beam in dark gray, the noncommon path of the local beam in light blue, and the noncommon path of the outgoing beam in green. }\label{fig:pseudorange-definition-and-TDI-toy-model}
\end{figure*}
\par
Each SC contains two movable optical sub-assemblies (MOSAs) connected by an optical fibre (see \cref{fig:LISA-Notation} for the labeling conventions). Each MOSA has an associated laser and houses a telescope, a free-falling test mass marking the end of the corresponding optical link, and an optical bench with three interferometers: the interspacecraft interferometer (ISI), in which the gravitational-wave signals eventually appear, the reference interfero\-meter (RFI) to compare local and adjacent lasers, and the test-mass interferometer (TMI) to sense the optical bench motion with respect to the free-falling test mass in direction of the optical link. The \si{\mega\hertz} beatnotes in these interferometers are detected with quadrant-photo-receivers (QPRs). They are digitized in analog-to-digital converters (ADCs) driven by the PMCs. Phasemeters extract the beatnote phases\footnote{Within the phasemeter both phase and frequency exist numerically. In the current design, the phasemeters deliver the beatnote frequencies with occasional phase anchor points.} using digital phase-locked loops (DPLLs). These phases are then downsampled to \SI{4}{\hertz} in a multi-stage decimation procedure (DEC) and telemetered to Earth (see \cref{fig:LISA-metrology-system}).
\subsection{The pseudorange and on-board delays}\label{sec:pseudorange-definition}
The pseudorange, denoted by $R_{ij}^{\hat{\tau}_i}$, is commonly defined as the difference between the SCET of the receiving SC at the event of reception and the SCET of the emitting SC at the event of emission \cite{hartwig2021instrumental}. It represents a combination of the light travel time between the emission at SC $j$ and the reception at SC $i$, and the differential SCET offset (see \cref{eq-ap:Pseudorange-TCB}). However, considering the complexity of the LISA metrology system, this definition appears to be rather vague: to what exactly do we relate the events of emission and reception? Two specifications are required here: we need to locate emission and reception, and we need to define the actual events.\par
It is convenient to consider emission and reception at the respective polarizing beam splitters (PBSs) in front of the telescopes (denoted PBS1 in \cite{BrzozowskiLISA-OB}), and to treat the on-board signal propa\-gation and processing on both SC as on-board delays (see \cref{fig:pseudorange-definition-and-TDI-toy-model}). Thus, we clearly separate the pseudorange from on-board delays. Note that this definition is not unique, the events of emission and reception could be located elsewhere, assuming that the on-board delays are defined accordingly.\par
The LISA optical links do not involve delta-pulse-like events. In order to define the actual events of emission and reception we, instead, use the instants when the light phase changes at the beginning of the first PRN code chip. At first glance, the PRN code might seem unfavorable for the pseudorange definition, as PRN and carrier phase are oppositely affected by the solar wind: the PRN phase is delayed by the group-delay, while the carrier phase is advanced by the phase delay. However, these effects are at the order of \SI{10}{\pico\m} (see \cref{appendix:Solar-Wind}), whereas our best pseudorange estimates are at \SI{0.1}{\milli\m} accuracy. Consequently, the solar wind dispersion can be neglected in the pseudorange definition.\par
When expressing the interferometric measurements according to this specified pseudorange definition, we need to consider the excluded on-board signal propagation and processing. For that purpose, we introduce two kinds of delay operators by their action on a function $f^{\hat{\tau}_{j}}$. The on-board delay operator describes delays due to on-board signal propagation and processing and is defined on the same SCET as the function it is acting on:
\begin{align}
	\textbf{D}_{x}^{\hat{\tau}_{j}}\:f^{\hat{\tau}_{j}}(\tau) &= f^{\hat{\tau}_{j}}\left(\tau - d_x^{\hat{\tau}_{j}}(\tau)\right).
\end{align}
$x$ is a place holder for any on-board delay, e.g., $\textbf{D}_{\text{qpr $\leftarrow$ pbs}}$ denotes the optical path length from the PBS to the QPR and $\textbf{D}_{\text{dec}}$ the decimation filter group delay. The interspacecraft delay operator is defined on a different SCET than the function it is acting on and applies the pseudorange as delay:
\begin{align}\label{eq:Inter-SC-Delay-Operator}
	\textbf{D}_{ij}^{\hat{\tau}_{i}}\:f^{\hat{\tau}_{j}}(\tau) &= f^{\hat{\tau}_{j}}\left(\tau - R_{ij}^{\hat{\tau}_{i}}(\tau)\right).
\end{align}
For on-board delays that differ between carrier, PRN, and sideband signals, we add the superscripts \emph{car}, \emph{prn}, and \emph{sb}, respectively. To trace the full path of a signal from the distant SC, we need to combine the interspacecraft delay operator for the interspacecraft signal propagation and the SCET conversion (considered at the PBS of the receiving SC) with on-board delay operators on both SC. The application of a delay operator to another time-dependent delay operator results in nested delays:
\begin{align}\label{eq:Nested-Delays}
	\textbf{D}_{x}^{\hat{\tau}_{i}}\:\textbf{D}_{ij}^{\hat{\tau}_{i}} f^{\hat{\tau}_{j}}(\tau) = f^{\hat{\tau}_{j}}\left(\tau - d_x^{\hat{\tau}_{i}}(\tau) - R_{ij}^{\hat{\tau}_{i}}\left(\tau - d_x^{\hat{\tau}_{i}}(\tau)\right)\right).
\end{align}
For a constant delay operator $\textbf{D}_{x}$ we can define the associated advancement operator $\textbf{A}_{x}$ acting as its inverse:
\begin{align}
	\textbf{A}^{\hat{\tau}_{j}}_{x}\: f^{\hat{\tau}_{j}}(\tau) &= f^{\hat{\tau}_{j}}\left(\tau + d_x^{\hat{\tau}_{j}}\right), \\
	\textbf{A}_{x} \:\textbf{D}_{x}\:f^{\hat{\tau}_{j}}(\tau)&= f^{\hat{\tau}_{j}}\left(\tau-d_x^{\hat{\tau}_{j}}+d_x^{\hat{\tau}_{j}}\right) = f^{\hat{\tau}_{j}}(\tau).
\end{align}
For advancement operators associated to propagation delays we write
\begin{align}
	\textbf{D}_{\text{qpr $\leftarrow$ pbs}}^{-1} = \textbf{A}_{\text{pbs $\leftarrow$ qpr}},
\end{align}
the subscript underlines that the advancement operator undoes the signal propagation. Below, we consider on-board delays as constant or slowly time varying so that their associated advancement operators are well-defined.
\subsection{Delays for TDI}\label{sec:delays-for-TDI}
In \cite{Hartwig:TDIwoSync} the pseudoranges are identified as the delays that need to be applied to cancel the laser noise in the alternative TDI topology (see \cref{fig:tdi-topologies}). Does this statement hold in the presence of on-board delays considering the refined pseudorange definition of \cref{sec:pseudorange-definition}? To identify the delays $\mathcal{D}_{ij}^{\hat{\tau}_i}$ that are required in TDI combinations to suppress the laser noise, let us set up a simple TDI toy model: we consider the 2 MOSAs depicted in \cref{fig:pseudorange-definition-and-TDI-toy-model} and the TDI combination, where we combine $\text{ISI}_{21}^{\hat{\tau}_2}$ with $\text{ISI}_{12}^{\hat{\tau}_1}$ delayed by $\mathcal{D}_{21}^{\hat{\tau}_2}$. We will identify the expression of $\mathcal{D}_{21}^{\hat{\tau}_2}$, which leads to a suppression of the noise from laser 12.\par
Let us first define the delays that are at play in this situation and highlighted in  \cref{fig:pseudorange-definition-and-TDI-toy-model}. We denote the delay that is common to both the local and the outgoing beam by $\textbf{D}_{\text{A}}^{\hat{\tau}_1}$. It corresponds to the path from the laser source to the beam splitter (BS), which devides them (denoted BS2 in \cite{BrzozowskiLISA-OB}). The delays associated to the paths only traversed by the local and the outgoing beam are called $\textbf{D}_{\text{B}}^{\hat{\tau}_1}$ and $\textbf{D}_{\text{C}}^{\hat{\tau}_2}$. Note that they represent combinations of the delays from the BS to the recombining beam splitters at the respective interspacecraft interferometers (ISI BSs) (denoted BS12 in \cite{BrzozowskiLISA-OB}), where the measurements are formed, with delays from the ISI BSs to the decimation filters, where the measurements are timestamped. Moreover, $\textbf{D}_{\text{C}}^{\hat{\tau}_2}$ combines an interspacecraft delay opera\-tor with on-board delays on both SC. Accordingly, we define the delays $\textbf{D}_{\text{A}}^{\hat{\tau}_2}$, $\textbf{D}_{\text{B}}^{\hat{\tau}_2}$, and $\textbf{D}_{\text{C}}^{\hat{\tau}_1}$ for laser 21.\par
Now the TDI delay $\mathcal{D}_{21}^{\hat{\tau}_2}$ can be derived as:
\begin{align}
	&\:\mathcal{D}_{21}^{\hat{\tau}_2} \: \text{ISI}_{12}^{\hat{\tau}_1}(\tau) + 	\text{ISI}_{21}^{\hat{\tau}_2}(\tau)\\
	=&\:\mathcal{D}_{21}^{\hat{\tau}_2} \left(
		\textbf{D}_{\text{C}}^{\hat{\tau}_1} \:	\textbf{D}_{\text{A}}^{\hat{\tau}_2} \: \Phi_{21}^{\hat{\tau}_2}(\tau) -
		\textbf{D}_{\text{B}}^{\hat{\tau}_1} \: \textbf{D}_{\text{A}}^{\hat{\tau}_1} \Phi_{12}^{\hat{\tau}_1}(\tau)
		\right)\nonumber\\
	+&\:\textbf{D}_{\text{C}}^{\hat{\tau}_2} \:	\textbf{D}_{\text{A}}^{\hat{\tau}_1} \: \Phi_{12}^{\hat{\tau}_1}(\tau) -
	\textbf{D}_{\text{B}}^{\hat{\tau}_2} \: \textbf{D}_{\text{A}}^{\hat{\tau}_2} \Phi_{21}^{\hat{\tau}_2}(\tau)\\
	=&\:\left( \textbf{D}_{\text{C}}^{\hat{\tau}_2} -  \mathcal{D}_{21}^{\hat{\tau}_2}\:\textbf{D}_{\text{B}}^{\hat{\tau}_1} \: \right) 	\textbf{D}_{\text{A}}^{\hat{\tau}_1} \: \Phi_{12}^{\hat{\tau}_1}(\tau)+ (...)\:\Phi_{21}^{\hat{\tau}_2}(\tau),
\end{align}
$\Phi_{12}^{\hat{\tau}_1}$ denotes the phase of laser 12. In the second step we factor out the common delay $\textbf{D}_{\text{A}}^{\hat{\tau}_1}$, which does not contribute to the TDI delay $\mathcal{D}_{21}^{\hat{\tau}_2}$. We ignore the phase $\Phi_{21}^{\hat{\tau}_2}$ of laser 21, as we focus on canceling the noise of laser 12. For that purpose we need to choose the delay $\mathcal{D}_{21}^{\hat{\tau}_2}$, such that the first bracket vanishes, i.e.
\begin{align}
	\mathcal{D}_{21}^{\hat{\tau}_2} =& \: \textbf{D}_{\text{C}}^{\hat{\tau}_2}\: (\textbf{D}^{-1}_{\text{B}})^{\hat{\tau}_1} = \textbf{D}_{\text{C}}^{\hat{\tau}_2}\: \textbf{A}_{\text{B}}^{\hat{\tau}_1}\\
	=&\:
	\textbf{D}_{\text{dec $\leftarrow$ qpr}}^{\hat{\tau}_2}\:\textbf{D}_{\text{qpr $\leftarrow$ pbs}}^{\hat{\tau}_{2}}\:\textbf{D}_{21}^{\hat{\tau}_{2}}\:\textbf{D}_{\text{pbs $\leftarrow$ bs}}^{\hat{\tau}_1}\nonumber\\
	&\:\textbf{A}_{\text{bs $\leftarrow$ qpr}}^{\hat{\tau}_1}\:\textbf{A}_{\text{qpr $\leftarrow$ dec}}^{\hat{\tau}_1}.\label{eq:TDI-delay}
\end{align}
The advancement operators are associated to the noncommon path of the local beam, the delay operators to the noncommon path of the outgoing beam. Hence, to cancel the noise of laser 12 we need to consider the difference between the delays applied to laser 12 in the ISI measurements on SC 1 and SC 2.\par 
The TDI delays $\mathcal{D}_{ij}^{\hat{\tau}_i}$ represent combinations of interspacecraft delay operators (pseudoranges) with on-board delay and advancement operators on both SC. We propose to calibrate all required on-board delays on ground before mission start and to measure the pseudorange during operation. The next section covers the four pseudo\-range observables. Before, we close this section with a few remarks on the required on-board delays. $\textbf{D}_{\text{pbs $\leftarrow$ bs}}^{\hat{\tau}_1}$ and $\textbf{D}_{\text{qpr $\leftarrow$ pbs}}^{\hat{\tau}_2}$ are small optical path length delays of the outgoing beam before transmission and after reception at the distant SC. $\textbf{A}_{\text{bs $\leftarrow$ qpr}}^{\hat{\tau}_1}$ is a small optical path length advancement of the local beam. These optical path lengths are in the order of \SI{10}{\centi\m} to \SI{1}{\m} \cite{BrzozowskiLISA-OB}.
$\textbf{D}_{\text{dec $\leftarrow$ qpr}}^{\hat{\tau}_2}$  is the signal processing delay on the receiving SC. $\textbf{A}_{\text{qpr $\leftarrow$ dec}}^{\hat{\tau}_1}$ is the corresponding advancement on the local SC. It can be decomposed into
\begin{align}
	\textbf{D}_{\text{dec $\leftarrow$ qpr}} =&\:\textbf{D}^{\text{car}}_{\text{dec}}\:\textbf{D}^{\text{car}}_{\text{dpll}}\:\textbf{D}_{\text{dpll $\leftarrow$ abee}}\:\textbf{D}^{\text{car}}_{\text{abee}} \nonumber\\
	&\:\textbf{D}_{\text{abee $\leftarrow$ qpr}}\:\textbf{D}^{\text{car}}_{\text{qpr}}.
\end{align}
The group delays of the quadrant-photo-receiver $\textbf{D}^{\text{car}}_{\text{qpr}}$ and the analog backend electronics $\textbf{D}^{\text{car}}_{\text{abee}}$ depend amongst others on the beatnote frequency \cite{BarrancoQPD}. Hence, they change over time and differ between carrier, sideband, and PRN signals. Together with the cable delays $\textbf{D}_{\text{abee$ \leftarrow$ qpr}}$ and $\textbf{D}_{\text{dpll$ \leftarrow$ abee}}$ they can amount to \SI{10}{\m}. The DPLL delay $\textbf{D}^{\text{car}}_{\text{dpll}}$ depends on the time-dependent beatnote amplitude \cite{Gerberding:Phasemeter, sutton2010laser}. All time-dependent contributions should be calibrated for all combinations of the time-dependent parameters. Hence, during operation they can be constructed with the help of SC monitors providing the corresponding parameter values, e.g., beatnote frequency and amplitude.

\section{Ranging measurements}\label{sec:ranging-measurements}
\begin{figure*}
	\begin{center}
		\includegraphics[width=1\textwidth]{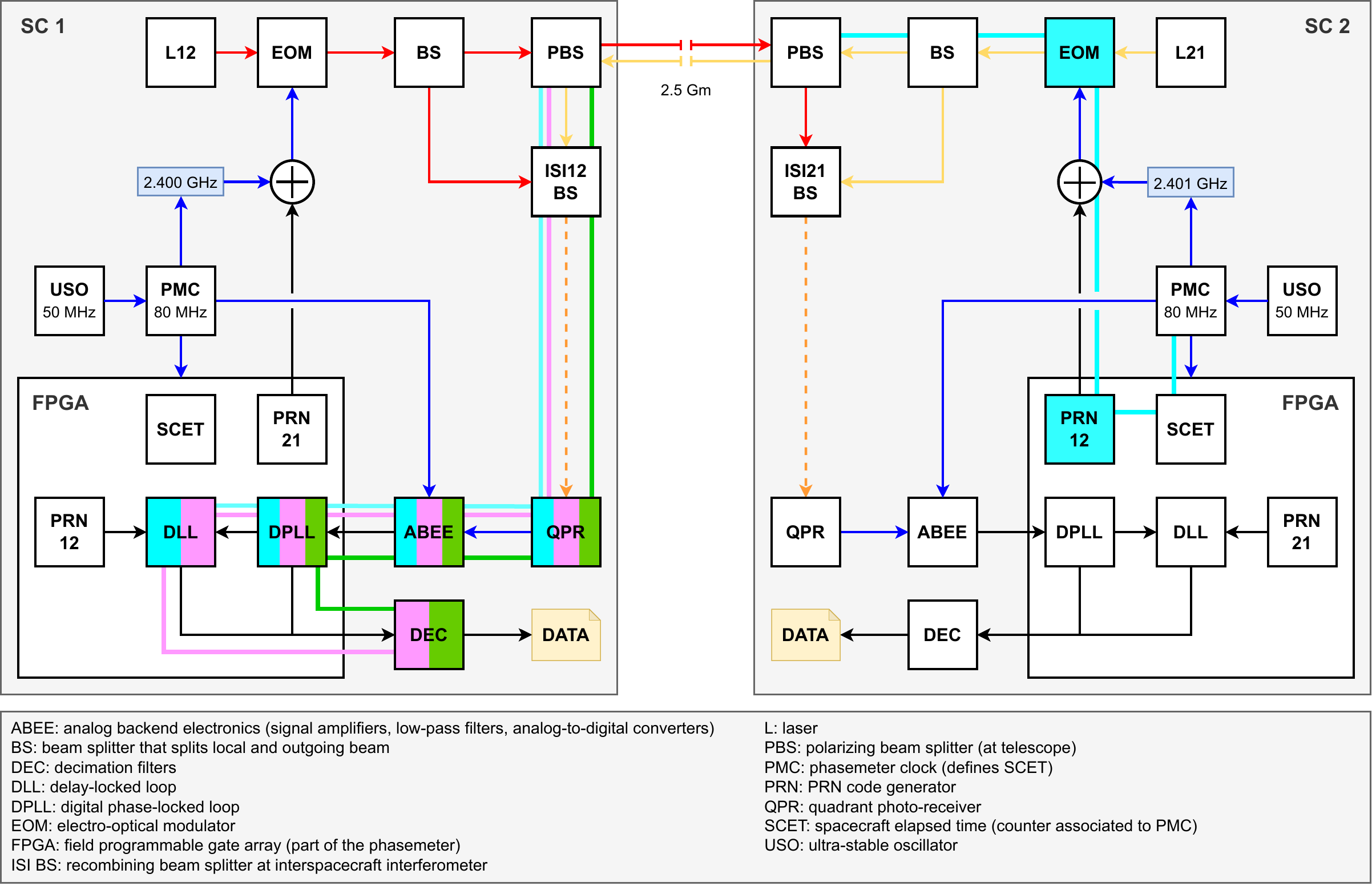}
	\end{center}
	\caption{We revisit \cref{fig:pseudorange-definition-and-TDI-toy-model}. Again we trace the lasers 12 and 21 to the ISI BSs, where they interfere and form beatnotes. The carriers are phase-modulated with the \si{\giga\hertz} clock and the PRN signals (follow the arrows from the PMC and the PRN to the EOM). Blue arrows indicate analog signals, black arrows digital ones. We show the USO frequency distribution (follow the arrows after the USOs) and illustrate the on-board signal processing (follow the arrows after the QPRs). We highlight the timestamping delays and offsets in the pseudorange observables caused by on-board delays. In light blue, we mark the PRNR offset from the pseudorange. The PRNR timestamping delay is drawn pink, the sideband timestamping delay is marked green.}\label{fig:LISA-metrology-system}
\end{figure*}
\subsection{PRN ranging (PRNR)}\label{sec:PRN-Ranging}
PRNR is the on-board ranging scheme in LISA. A set of 6 PRN sequences has been computed such that the cross-correlations and the auto-correlations for nonzero delays are minimized. These PRN codes are associated to the 6 optical links in the LISA constellation. The PRN codes are genera\-ted according to the respective PMCs and imprinted onto the laser beams by phase-modulating the carriers in electro-optical modulators (EOMs) (see \cref{fig:LISA-metrology-system}). In each phasemeter, DPLLs are applied to extract the beatnote phases. The PRN codes show up in the DPLL error signals since the DPLL bandwidth of \SI{10}{\kilo\Hz} to \SI{100}{\kilo\Hz} is lower than the PRN chipping rate of about \SI{1}{\mega\Hz}. In a delay-locked loop (DLL), these error signals are correlated with PRN codes generated according to the local SCET. The local delay that maximizes the correlation yields a pseudorange measurement: the PRNR \cite{EstebanPRN1, EstebanPRN2}.\par
We now derive the PRNR observation equation carefully taking into account on-board delays. We model the path of the PRN code from the distant SC to the local DLL by applying delay operators to the distant SCET:
\begin{align}\label{eq:Distant-SCET-Delayed-to-Local-DLL}
	\textbf{D}_{\text{dll $\leftarrow$ pbs}}^{\text{prn},\:\hat{\tau}_{i}} \: \textbf{D}_{ij}^{\hat{\tau}_{i}} \: \textbf{D}_{\text{pbs $\leftarrow$ pmc}}^{\text{prn},\:\hat{\tau}_{j}}\:\hat{\tau}_{j}^{\hat{\tau}_{j}}(\tau).
\end{align}
The two on-board delays can be decomposed into
\begin{align}
	\textbf{D}_{\text{pbs $\leftarrow$ pmc}}^{\text{prn}}=\:&\textbf{D}_{\text{pbs $\leftarrow$ eom}}\:\textbf{D}_{\text{eom $\leftarrow$ prn}}\nonumber\\
	&\textbf{D}_{\text{prn}}\:\textbf{D}_{\text{prn $\leftarrow$ pmc}},\\
	\textbf{D}_{\text{dll $\leftarrow$ pbs}}^{\text{prn}}=\:&\textbf{D}_{\text{dll}}\:\textbf{D}_{\text{dpll}}^{\text{prn}}\:\textbf{D}_{\text{dpll $\leftarrow$ abee}}\:\textbf{D}_{\text{abee}}^{\text{prn}}\:\textbf{D}_{\text{abee $\leftarrow$ qpr}}\nonumber\\
	&\textbf{D}_{\text{qpr}}^{\text{prn}}\:\textbf{D}_{\text{qpr $\leftarrow$ pbs}}.
\end{align}
$\textbf{D}_{\text{pbs $\leftarrow$ pmc}}^{\text{prn}}$ consists of the cable delays from the PMC to the EOM, the processing delay due to the PRN code generation, and the optical path length from the EOM to the PBS. All these delays are constant at the sensitive scale of PRNR, so that we do not have to consider delay nesting in $\textbf{D}_{\text{pbs $\leftarrow$ pmc}}^{\text{prn}}$. We added the superscript \emph{prn} because this path is different for the sideband signal (see \cref{fig:LISA-metrology-system}). $\textbf{D}_{\text{dll $\leftarrow$ pbs}}^{\text{prn}}$ is explained in the next paragraph as part of the PRNR timestamping delay. At the DLL, the received PRN codes are correlated with identical codes generated according to the local SCET. We model this correlation as the difference between the local SCET and the delayed distant SCET (\cref{eq:Distant-SCET-Delayed-to-Local-DLL}), and we apply $\textbf{D}^{\text{prn}}_{\text{dec}}$ to model the group delay of the decimation filters applicable to PRN ranging:
\begin{align}\label{eq:PRNR-Expressed-with-All-Delay-Operators}
	\textbf{D}^{\text{prn}}_{\text{dec}}\left(\hat{\tau}_{i}^{\hat{\tau}_{i}}(\tau) - \textbf{D}_{\text{dll $\leftarrow$ pbs}}^{\text{prn},\:\hat{\tau}_{i}} \: \textbf{D}_{ij}^{\hat{\tau}_{i}} \: \textbf{D}_{\text{pbs $\leftarrow$ pmc}}^{\text{prn},\:\hat{\tau}_{j}}\:\hat{\tau}_{j}^{\hat{\tau}_{j}}(\tau)\right).
\end{align}
To see how the on-board delays affect the PRNR we expand \cref{eq:PRNR-Expressed-with-All-Delay-Operators} applying \cref{eq:SCET-in-SCET}:
\begin{align}
	&\:\textbf{D}^{\text{prn}}_{\text{dec}}\Big{(}\hat{\tau}_{i}^{\hat{\tau}_{i}}(\tau) -\hat{\tau}_{j}^{\hat{\tau}_{j}}\big(\tau - d^{\hat{\tau}_{i}}_{\text{dll $\leftarrow$ pbs}}\nonumber\\
	&\hspace{3.22cm} - R_{ij}^{\hat{\tau}_{i}}\big(\tau - d^{\hat{\tau}_{i}}_{\text{dll $\leftarrow$ pbs}}\big)\nonumber\\
	&\hspace{3.22cm} - d^{\hat{\tau}_{j}}_{\text{pbs $\leftarrow$ pmc}}\big{)}\Big{)}\nonumber\\
	=&\:\textbf{D}_{\text{dec $\leftarrow$ pbs}}^{\text{prn},\:\hat{\tau}_{i}}\:R_{ij}^{\hat{\tau}_{i}}\left(\tau\right)
	+O^{\text{prn}}_{ij}.
\end{align}
The on-board delays cause the PRNR timestamping delay $\textbf{D}_{\text{dec $\leftarrow$ pbs}}^{\text{prn}}$ and the PRNR offset $O^{\text{prn}}_{ij}$:
\begin{align}
	\textbf{D}_{\text{dec $\leftarrow$ pbs}}^{\text{prn}}=&\:\textbf{D}^{\text{prn}}_{\text{dec}}\:\textbf{D}_{\text{dll}}\:\textbf{D}_{\text{dpll}}^{\text{prn}}\:\textbf{D}_{\text{dpll $\leftarrow$ abee}}\:\textbf{D}_{\text{abee}}^{\text{prn}}\nonumber\\&\:\textbf{D}_{\text{abee $\leftarrow$ qpr}}\:\textbf{D}_{\text{qpr}}^{\text{prn}}\:\textbf{D}_{\text{qpr $\leftarrow$ pbs}},\label{eq:PRN-Delay}\\
	O_{ij}^{\text{prn}}=&\:d^{\hat{\tau}_{i}}_{\text{dll $\leftarrow$ pbs}}+d^{\hat{\tau}_{j}}_{\text{pbs $\leftarrow$ pmc}}.\label{eq:PRNR-Offset}
\end{align}
The PRNR timestamping delay has similar constituents as the on-board delays appearing in the TDI delay $\mathcal{D}_{ij}$, they are marked pink in \cref{fig:LISA-metrology-system}. However, most of them are frequency or amplitude dependent. Therefore, they differ between carrier and PRN signals. As for the TDI delay, we propose to individually calibrate all constituents of the PRNR time\-stamping delay on ground before mission start. Hence, during operation $\textbf{D}_{\text{dec $\leftarrow$ pbs}}^{\text{prn}}$ can be compensated in an initial data treatment by application of its associated advancement operator $\textbf{A}_{\text{pbs $\leftarrow$ dec}}^{\text{prn}}$. After that, the PRNR observation equation including ranging noise and PRN ambiguity can be written as:
\begin{align}
	\textbf{A}_{\text{pbs $\leftarrow$ dec}}^{\text{prn},\:\hat{\tau}_{i}}\:\text{PRNR}_{ij}^{\hat{\tau}_{i}}(\tau) =&\: R_{ij}^{\hat{\tau}_{i}}(\tau) + O^{\text{prn}}_{ij} + N^{\text{prn}}_{ij}(\tau)\nonumber\\
	-&\:a^{\text{prn}}_{ij}(\tau) \cdot l.
\end{align}
$l$ denotes the finite PRN code length. We use \SI{400}{\kilo\m} as a placeholder, the final value has not been decided. The finite PRN code length leads to an ambiguity, $a^{\text{prn}}_{ij}$ denote the associated ambiguity integers \cite{EstebanPRN2}. $N^{\text{prn}}_{ij}$ is the white ranging noise with an rms amplitude of about \SI{0.3}{\m} at the \SI{4}{\hertz} data rate, which is due to shot noise and PRN code interfe\-rence \cite{Heinzel-Ranging, sutton2010laser}. The latter refers to the interference between the PRN code modulated onto the received laser at the distant SC and the other code modulated onto the local laser at the local SC. The PRNR offset $O^{\text{prn}}_{ij}$ involves contributions on the emitter and on the receiver side (see \cref{eq:PRNR-Offset}), they are marked light blue in \cref{fig:LISA-metrology-system}. It can amount to \SI{10}{\m} and more \cite{sutton2010laser, Euringer-Code-Tracking}. Similar to the PRNR timestam\-ping delay, we propose to calibrate the PRNR offset on ground, so that it can be subtracted in an initial data treatment.
\subsection{Sideband ranging (SBR)}\label{sec:SBR}
For the purpose of in-band clock noise correction in the INReP, a clock noise transfer between the SC is implemented \cite{Heinzel-Ranging}: the \SI{80}{\mega\hertz} PMC signals are upconverted to $\nu^{\text{m}}_{l}=\SI{2.400}{\giga\hertz}$ and $\nu^{\text{m}}_{r}=\SI{2.401}{\giga\hertz}$ for left and right-handed MOSAs, respectively (see \cref{fig:LISA-Notation} for the defi\-nition of left and right-handed MOSAs). The EOMs phase-modulate the carriers with the upconverted PMC signals, thereby creating clock sidebands. We show below that the beatnotes between these clock sidebands constitute a pseudorange observable.\par
Considering on-board delays, the difference between carrier and sideband beatnotes can be written as
\begin{align}\label{eq:Sideband-Beatnote}
	\text{ISI}&_{ij}^{\hat{\tau}_i}(\tau) - \text{ISI}_{\text{sb},\:ij}^{\hat{\tau}_i}(\tau) = -\:\textbf{D}^{\text{sb},\:\hat{\tau}_i}_{\text{dec $\leftarrow$ isi bs}}\nonumber\\
	\Big\{&\textbf{D}^{\hat{\tau}_i}_{\text{isi bs $\leftarrow$ pbs}}\: \textbf{D}^{\hat{\tau}_i}_{ij}\left(\textbf{D}^{\text{sb},\:\hat{\tau}_j}_{\text{pbs $\leftarrow$ pmc}}\:\nu^{\text{m}}_{ji}\:\hat{\tau}_j^{\hat{\tau}_j}(\tau) + \nu^{\text{m}}_{ji}\:M^{\hat{\tau}_j}_{ji}(\tau)\right)\nonumber\\
	-\Big(&\textbf{D}^{\text{sb},\:\hat{\tau}_i}_{\text{isi bs $\leftarrow$ pmc}}\:\nu^{\text{m}}_{ij}\:\hat{\tau}_i^{\hat{\tau}_i}(\tau) + \nu^{\text{m}}_{ij}\:M^{\hat{\tau}_i}_{ij}(\tau) \Big) \Big\}.
\end{align}
$\textbf{D}^{\text{sb}}_{\text{pbs $\leftarrow$ pmc}}$ and $\textbf{D}^{\text{sb}}_{\text{isi bs $\leftarrow$ pmc}}$ are the delay operators associated to the paths from the PMC to the PBS and from the PMC to the ISI BS. They can be decomposed into:
\begin{align}
	\textbf{D}^{\text{sb}}_{\text{(p|isi)bs$\leftarrow$pmc}}=\textbf{D}_{\text{(p|isi)bs$\leftarrow$eom}}\:\textbf{D}_{\text{eom$\leftarrow$pmc}}\:\textbf{D}_{\text{up}}.
\end{align}
$\textbf{D}_{\text{up}}$ is the upconversion delay due to phase-locking a \SI{2.40(1)}{\giga \hertz} oscillator to the \SI{80}{\mega\hertz} PMC signal, $\textbf{D}_{\text{eom $\leftarrow$ pmc}}$ is the cable delay from the PMC to the EOM.
$\nu^{\text{m}}_{ij}$ is the upconverted USO frequency associated to $\text{MOSA}_{ij}$. Since \cref{eq:Sideband-Beatnote} is expressed in the SCET, all clock imperfections are included into $\hat{\tau}_i^{\hat{\tau}_i}(\tau)$. The modulation noise $M^{\hat{\tau}_i}_{ij}$ contains any additional jitter collected on the path $\textbf{D}^{\text{sb}}_{\text{(p|isi)bs $\leftarrow$ pmc}}$, e.g., due to the electrical frequency upconverters. The amplitude spectral densities (ASDs) of the modulation noise for left and right-handed MOSAs are specified to be below \cite{Hartwig:TDIwoSync, Barke2015Thesis}
\begin{align}\label{eq:right-handed-Mod-ASD}
	\sqrt{S_{M_l}(f)} &= 2.5 \times 10^{-6}\si{\m\hertz\tothe{-0.5}}\left(\frac{f}{\si{Hz}}\right)^{-2/3},\\
	\sqrt{S_{M_r}(f)} &= 2.5 \times 10^{-5}\si{\m\hertz\tothe{-0.5}}\left(\frac{f}{\si{Hz}}\right)^{-2/3}.
\end{align}
The modulation noise on left-handed MOSAs is one order of magnitude lower, because the pilot tone for the ADC jitter correction (it is the ultimate phase reference) is derived from the \SI{2.400}{\giga\hertz} clock signal.
\par
To derive a pseudorange observation equation from the sideband beatnote we expand \cref{eq:Sideband-Beatnote} using \cref{eq:SCET-in-SCET}. We apply $\textbf{A}^{\text{sb}}_{\text{pbs $\leftarrow$ dec}}$ to avoid nested delays in the pseudorange:
\begin{align}\label{eq:Sideband-Beatnote-Advanced}
	&\hspace{0.6cm}\textbf{A}^{\text{sb},\:\hat{\tau}_i}_{\text{pbs $\leftarrow$ dec}}\left( \text{ISI}_{ij}^{\hat{\tau}_i}(\tau) - \text{ISI}_{\text{sb},\:ij}^{\hat{\tau}_i}(\tau) \right)\nonumber\\
	=&-\nu^{\text{m}}_{ji}\:\textbf{D}^{\hat{\tau}_i}_{ij}\left( \textbf{D}^{\text{sb},\:\hat{\tau}_j}_{\text{pbs $\leftarrow$ pmc}}\:\hat{\tau}_j^{\hat{\tau}_j}(\tau) + M^{\hat{\tau}_j}_{ji}(\tau)\right)\nonumber\\
	+&\:\nu^{\text{m}}_{ij}\:\textbf{A}^{\hat{\tau}_i}_{\text{pbs $\leftarrow$ isi bs}}\left(\textbf{D}^{\text{sb},\:\hat{\tau}_i}_{\text{isi bs $\leftarrow$ pmc}}\:\hat{\tau}_i^{\hat{\tau}_i}(\tau) + M^{\hat{\tau}_i}_{ij}\right)\nonumber\\
	=&\:\left(\nu^{\text{m}}_{ij} - \nu^{\text{m}}_{ji}\right)\tau + \nu^{\text{m}}_{ji}\:R_{ij}^{\hat{\tau}_{i}}(\tau)\nonumber\\
	+&\:\nu^{\text{m}}_{ji}\cdot d^{\hat{\tau}_j}_{\text{pbs $\leftarrow$ pmc}} -\nu^{\text{m}}_{ij}\cdot \left(d^{\hat{\tau}_i}_{\text{isi bs $\leftarrow$ pmc}} - d^{\hat{\tau}_i}_{\text{pbs $\leftarrow$ isi bs}}\right)\nonumber\\
	+&\:\nu^{\text{m}}_{ij}\:\textbf{A}^{\hat{\tau}_i}_{\text{pbs $\leftarrow$ isi bs}}\:M^{\hat{\tau}_i}_{ij}(\tau)-\nu^{\text{m}}_{ji}\:\textbf{D}^{\hat{\tau}_i}_{ij} M^{\hat{\tau}_j}_{ji}(\tau).
\end{align}
We subtract the \SI{1}{\mega\hertz} ramp and then refer to \cref{eq:Sideband-Beatnote-Advanced} as sideband ranging (SBR). Taking into account that the SBR phase is defined up to a cycle, the SBR can be written as
\begin{align}\label{eq:Sideband-Ranging}
	\text{SBR}_{ij}^{\hat{\tau}_i}(\tau) &= \textbf{A}^{\text{sb},\:\hat{\tau}_i}_{\text{pbs $\leftarrow$ dec}}\left( \text{ISI}_{ij}^{\hat{\tau}_i}(\tau) - \text{ISI}_{\text{sb},\:ij}^{\hat{\tau}_i}(\tau) \right) \pm \SI{1}{\mega\hertz}\: \tau \nonumber\\ &=\nu^{\text{m}}_{ji}\:R_{ij}^{\hat{\tau}_{i}}(\tau)+ O^{\text{sb}}_{ij} + N^{\text{sb}}_{ij}(\tau) - a^{\text{sb}}_{ij}(\tau).
\end{align}
$a^{\text{sb}}_{ij}$ denote the SBR ambiguity integers. Expressed as length, the SBR ambiguity is \SI{12.5}{\centi\m} corresponding to the wavelength of the \si{\giga\hertz} sidebands. The SBR offset 
\begin{align}
	O^{\text{sb}}_{ij}=&\:\nu^{\text{m}}_{ji}\cdot d^{\hat{\tau}_j}_{\text{pbs $\leftarrow$ pmc}}\nonumber\\ -&\:\nu^{\text{m}}_{ij}\cdot \left(d^{\hat{\tau}_i}_{\text{isi bs $\leftarrow$ pmc}} - d^{\hat{\tau}_i}_{\text{pbs $\leftarrow$ isi bs}}\right)\label{eq:SBR-offset}
\end{align}
can be thought of as the differential phase accumulation of local and distant PMC signals on their paths to the respective PBSs. Similar to the PRNR offset the SBR offset could be measured on ground. $N^{\text{sb}}_{ij}$ denotes the appearance of the modulation noise in the SBR:
\begin{align}
	N^{\text{sb}}_{ij} =&\:\nu^{\text{m}}_{ij}\:\textbf{A}^{\hat{\tau}_i}_{\text{pbs $\leftarrow$ isi bs}}\:M^{\hat{\tau}_i}_{ij}(\tau)-\nu^{\text{m}}_{ji}\:\textbf{D}^{\hat{\tau}_i}_{ij} M^{\hat{\tau}_j}_{ji}(\tau).\label{eq:Modulation-Noise-Appearance}
\end{align}
This is a combination of left and right-handed modulation noise, their rms amplitudes are $2.9 \times 10^{-5}\:\si{\m}$ and 2.9 $\times 10^{-4}\:\si{\m}$, respectively. As shown in \cite{Hartwig:TDIwoSync}, it is possible to combine carrier and sideband beatnotes from the RFI to form measurements of the dominating right-handed modulation noise, which can, thus, be subtracted from the $\text{SBR}$s (see \cref{appendix:RH-modulation-noise}).
\par
The advancement operator $\textbf{A}_{\text{pbs $\leftarrow$ dec}}^{\text{sb}}$ (see \cref{eq:Sideband-Beatnote-Advanced}) is associated to the delay operator $\textbf{D}_{\text{dec $\leftarrow$ pbs}}^{\text{sb}}$, to which we refer as sideband timestamping delay. The sideband timestamping delay can be decomposed into:
\begin{align}\label{eq:Sideband-Delay}
	\textbf{D}_{\text{dec $\leftarrow$ pbs}}^{\text{sb}} =&\:\textbf{D}^{\text{sb}}_{\text{dec}}\:\textbf{D}^{\text{sb}}_{\text{dpll}}\:\textbf{D}_{\text{dpll $\leftarrow$ abee}}\:\textbf{D}^{\text{sb}}_{\text{abee}} \nonumber\\
	&\:\textbf{D}_{\text{abee $\leftarrow$ qpr}}\:\textbf{D}^{\text{sb}}_{\text{qpr}}\:\textbf{D}_{\text{qpr $\leftarrow$ pbs}},
\end{align}
these constituents are marked green in \cref{fig:LISA-metrology-system}. As for the PRNR timestamping delay, we propose to individually calibrate all its constituents on ground. The sideband timestamping delay can then be compensated in an initial data treatment by application of its associated advancement operator (see \cref{eq:Sideband-Beatnote-Advanced}).\par
In reality, the beatnotes are expected to be delivered not in phase, but in frequency with occasional phase anchor points. Therefore, we consider the derivative of \cref{eq:Sideband-Ranging}, we refer to it as sideband range rate ($\dot{\text{SBR}}$):
\begin{align}
	\dot{\text{SBR}}_{ij}^{\hat{\tau}_i}(\tau)&=\nu^{\text{m}}_{ji}\:\dot{R}_{ij}^{\hat{\tau}_{i}}(\tau) + \dot{N}^{\text{sb}}_{ij}(\tau).\label{eq:dotSBR}
\end{align}
The sideband range rates are an offset-free and unambiguous measurement of the pseudorange time derivatives. Phase anchor points enable their integration, so that we recover \cref{eq:Sideband-Ranging}.
\subsection{Time-delay interferometric ranging (TDIR)}\label{sec:TDIR}
TDI builds combinations of delayed ISI and RFI carrier beatnotes to virtually form equal-arm interferometers, in which laser frequency noise is suppressed. In the alternative TDI topology, the corresponding TDI delays $\mathcal{D}_{ij}$ are given by the pseudoranges in combination with on-board delays (see \cref{eq:TDI-delay}). Time-delay interferome\-tric ranging (TDIR) turns the main scientific data streams themselves into an absolute ranging observable: it mini\-mizes the power integral of the laser frequency noise in the TDI combinations by varying the delays that are applied to the beatnotes \cite{TintoTDIR}. When doing this before clock synchronization to TCB, i.e., with the beatnotes sampled according to the respective SCETs, the TDI delays $\mathcal{D}_{ij}$ show up at the very minimum of that integral. Thus, TDIR constitutes an unbiased observable of the TDI delays $\mathcal{D}_{ij}$, which only requires the interferometric measurements.\par
Below, we consider TDI in frequency \cite{Bayle:TDI-in-Frequency}. Therefore, 
we introduce the Doppler-delay operator associated to the TDI delay:
\begin{align}
	\dot{\mathcal{D}}^{\hat{\tau}_i}_{ij}\:f^{\hat{\tau}_j}(\tau)= \left(1 - \dot{R}_{ij}^{\hat{\tau}_i}(\tau)\right)\mathcal{D}^{\hat{\tau}_i}_{ij}\: f^{\hat{\tau}_j}\left(\tau\right).
\end{align}
Here we assume the on-board delay constituents of $\mathcal{D}^{\hat{\tau}_i}_{ij}$ to be slowly time-varying, so that only the pseudorange time derivative appears in the Doppler factor. We use the shorthand notation
\begin{align}
	\dot{\mathcal{D}}^{\hat{\tau}_i}_{ijk} = \dot{\mathcal{D}}^{\hat{\tau}_i}_{ij}\:\dot{\mathcal{D}}^{\hat{\tau}_j}_{jk}
\end{align}
to indicate chained Doppler-delay operators. In this paper we neglect on-board delays in the RFI beatnotes. We start our consideration of TDIR from the intermediary TDI variables $\eta_{ij}$. These are combinations of the ISI and RFI carrier beatnotes to eliminate the laser frequency noise contributions of right-handed lasers. In terms of the $\eta_{ij}$ the second-generation TDI Michelson variables can be expressed as \cite{Otto2015TDI}
\begin{align}\label{eq:TDI-Michelson-2}
	&X_2^{\hat{\tau}_1} =\left(1 - \dot{\mathcal{D}}^{\hat{\tau}_1}_{121} - \dot{\mathcal{D}}^{\hat{\tau}_1}_{12131} + \dot{\mathcal{D}}^{\hat{\tau}_1}_{1312121}\right)\left(\eta^{\hat{\tau}_1}_{13} - \dot{\mathcal{D}}^{\hat{\tau}_1}_{13}\eta^{\hat{\tau}_3}_{31}\right)\nonumber\\
	&-\left(1 - \dot{\mathcal{D}}^{\hat{\tau}_1}_{131} - \dot{\mathcal{D}}^{\hat{\tau}_1}_{13121} + \dot{\mathcal{D}}^{\hat{\tau}_1}_{1213131}\right)\left(\eta^{\hat{\tau}_1}_{12} - \dot{\mathcal{D}}^{\hat{\tau}_1}_{12}\eta^{\hat{\tau}_2}_{21}\right)
\end{align}
$Y_2^{\hat{\tau}_2}(\tau)$ and $Z_2^{\hat{\tau}_3}(\tau)$ are obtained by cyclic permutation of the indices. For later reference, we also state the first generation TDI Michelson variables:
\begin{align}\label{eq:TDI-Michelson-1}
	X_1^{\hat{\tau}_1} =&\:(1 - \dot{\mathcal{D}}^{\hat{\tau}_1}_{121})\left(\eta^{\hat{\tau}_1}_{13} - \dot{\mathcal{D}}^{\hat{\tau}_1}_{13}\eta^{\hat{\tau}_3}_{31}\right)\nonumber\\
	-&\:(1 - \dot{\mathcal{D}}^{\hat{\tau}_1}_{131})\left(\eta^{\hat{\tau}_1}_{12} - \dot{\mathcal{D}}^{\hat{\tau}_1}_{12}\eta^{\hat{\tau}_2}_{21}\right).
\end{align}
\par
In the framework of TDIR, the delays applied in TDI are parameterized by a model, e.g., by a polynomial model. We minimize the power integral of the TDI combinations by varying the model parameters. TDIR attempts to minimize the in-band laser frequency noise residual. Therefore, we apply a band-pass filter to first remove other contributions appearing out-of-band, i.e., slow drifts and contributions above 1Hz that are domi\-nated by aliasing and interpolation errors. We use the TDIR estimator
\begin{align}\label{eq:TDIR-integral}
	\text{TDIR}_{ij}^{\hat{\tau}_i} &= \min_{\Theta}\frac{1}{T}\int_{\frac{1}{T}}^{T} \left[\tilde{X}_2^{\hat{\tau}_1}\right]^2 + \left[\tilde{Y}_2^{\hat{\tau}_2}\right]^2 + \left[\tilde{Z}_2^{\hat{\tau}_3}\right]^2\:\text{d}t,
\end{align}
similar to the one proposed by \cite{TintoTDIR}.\footnote{Note that this is not the optimal TDIR estimator, as the noise shapes and the correlations between different channels are not taken into account.} $\Theta$ denotes the parameters of the delay model, the tilde indicates the filtered TDI combinations.\par
The TDIR accuracy, we denote it by $\sigma^{\text{tdir}}$, increases with the integration time $T$ (length of telemetry dataset). It is in the order of \cite{TintoTDIR}:
\begin{align}
	\sigma^{\text{tdir}}(T) \propto \SI{10}{\centi\m}\: \sqrt{\frac{\si{\day}}{T}},\label{eq:TDIR-accuracy}
\end{align}
where \si{\day} stands for day. Hence, we require about 1000s of data to reach meter accuracy. 
\subsection{Ground-observation based ranging (GOR)}\label{sec:GOR}
The mission operation center (MOC) provides orbit determinations (ODs) via the ESA tracking stations and MOC time correlations (MOC-TCs). When combined properly, these two on-ground measurements form a pseudorange observable referred to as ground-observation based ranging (GOR). It has an uncertainty of about $\SI{50}{\kilo \m}$ due to uncertainties in both the OD and the MOC-TC. Yet, it yields valuable information. It is unambiguous, hence it allows to resolve the PRNR ambiguities.\par 
The OD yields information about the absolute positions and velocities of the three SC. New orbit determinations are published every few days. For the position and velocity measurements in the line of sight, radial (with respect to the sun) and cross-track direction conservative estimations by ESA state the uncertainties as \SI{2}{\kilo\m} and \SI{4}{\milli\m \per \s}, \SI{10}{\kilo\m} and \SI{4}{\milli\m \per \s}, \SI{50}{\kilo\m} and \SI{5}{\centi\m \per \s}, respectively \cite{MartensOD}. The MOC-TC is a measurement of the SCET desynchronization from TCB. It is determined during the telemetry contacts via a comparison of the SCET associated to the emission of a telemetry packet and the TCB of its reception on Earth taking into account the down link delay. We expect the accuracy of the MOC-TC to be better than \SI{0.1}{\milli\s} (corresponds to \SI{30}{\kilo\m}). This uncertainty is due to unexact knowledge of the SC-to-ground-station separation, as well as inaccuracies in the time tagging process on board and on ground.\par 
As shown in \cref{appendix:pr_TCB}, the pseudorange can be expressed in TCB as function of the reception time:
\begin{align}\label{eq:pseudorange}
	R_{ij}^t(t)&= (1 + \delta \dot{\hat{\tau}}_{j}^{t}(t)) \cdot  d_{ij}^{t}(t) + \delta\hat{\tau}_{ij}^{t}(t).
\end{align}
$d_{ij}^{t}$ denotes the light travel time from SC $j$ to SC $i$, $\delta\hat{\tau}_{ij}^{t}$ the offset between the involved SCETs, and $\delta \dot{\hat{\tau}}_{j}^{t}$ the SCET drift of the emitting SC with respect to TCB. The light travel times can be expressed in terms of the ODs \cite{HeesLTT}:
\begin{align}
	d^t_{\text{od},\:ij}(t) &=\frac{1}{c} L^t_{ij}(t) + \frac{1}{c^2}\: \vec{L}^t_{ij}(t)\cdot \vec{v}_{j}^t(t) + O(c^{-3}),\\
	\vec{L}_{ij}&= \vec{r}_{i} - \vec{r}_{j}, \: L_{ij} = \vert \vec{L}_{ij} \vert.
\end{align}
$\vec{r}_{i}$ denotes the position of the receiving SC, $\vec{r}_{j}$ and $\vec{v}_{j}$ are position and velocity of the emitting one. The $O(c^{-3})$ terms contribute to the light travel time at the order of \SI{10}{\m}. They are negligible compared to the large uncertainties of the orbit determination. Combining these light travel time estimates with the MOC-TCs allows us to write the GOR as
\begin{align}
	\text{GOR}_{ij}^t(t) &= d^t_{\text{od},\:ij}(t)+\delta\hat{\tau}^t_{\text{tc}, \:ij}(t)=R_{ij}^t(t) + N^{\text{gor}}_{ij},\\
	\delta\hat{\tau}^t_{\text{tc}, \:ij}(t)&=\delta\hat{\tau}^t_{\text{tc}, \:i}(t) - \delta\hat{\tau}^t_{\text{tc}, \:j}(t).
\end{align}
$\delta\hat{\tau}^t_{\text{tc}, \:i}$ denotes the MOC-TC of SC $i$ and $N^{\text{gor}}\sim\SI{50}{\kilo \m}$ the GOR uncertainty. Note that the ODs and the MOC-TCs (hence, also the GOR) are given in TCB, while all other pseudorange observables are sampled in the respective SCETs. This desynchronization is negligible: the desynchronization can amount up to \SI{10}{\s} after the ten year mission time, the pseudoranges drift with \si{10} to \SI{100}{\m \s \tothe{-1}} (see central plot in \cref{fig:PRNR-Unwrapping}). Hence, neglecting the desynchronization leads to errors of about \si{100} to \SI{1000}{\m}, which are negligible compared to the large GOR uncertainty.

\section{Ranging sensor fusion}\label{sec:sensor-fusion}
\begin{figure*}
	\begin{center}
		\includegraphics[width=1\textwidth]{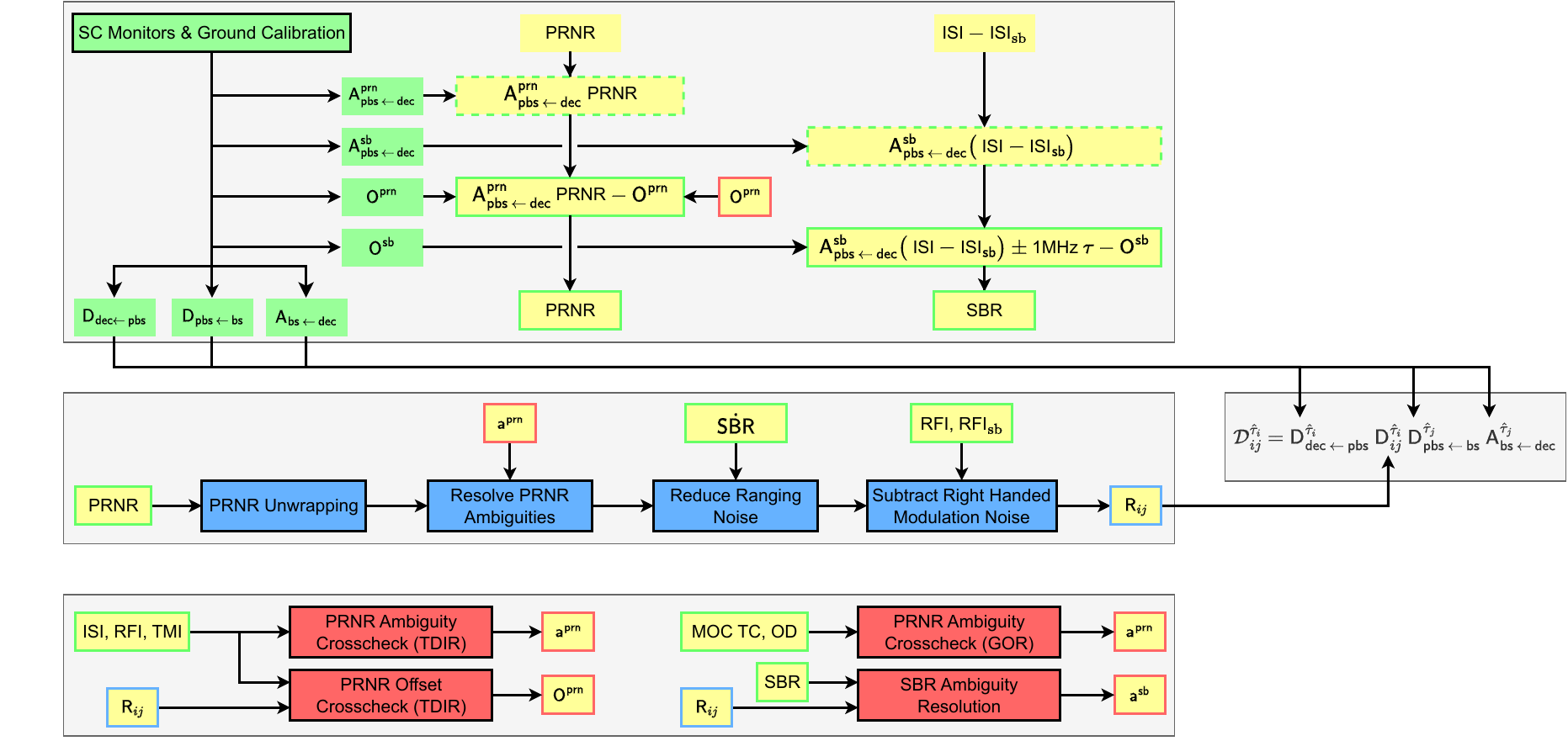}
	\end{center}
	\caption{We illustrate the three-stage ranging sensor fusion. Processing elements are drawn with a black frame. In the upper part we show the initial data treatment. Products of the on-ground calibration (the various delays and offsets) are drawn green. Raw datasets are drawn yellow, after the initial data treatment we add a green frame. In the central part we show the core ranging processing. Its output, the pseudoranges, are drawn with a blue frame. In the right box we show how the pseudoranges are combined with on-board delays to form the TDI delays. In the lower part we show simultaneous crosschecks of PRNR ambiguity, PRNR offset, and SBR ambiguity. Products of these crosschecks are drawn with a red frame.}\label{fig:sensor-fusion}
\end{figure*}
To combine the four pseudorange observables, we propose a three-stage ranging sensor fusion (RSF) consisting of an initial data treatment, a ranging processing, and crosschecks. The ranging processing (central part of \cref{fig:sensor-fusion}) refers to the ranging-related routines, which need to run continuously during operation. These are the PRNR unwrapping, and the reduction of ranging and right-handed modulation noise. Simultaneously, the PRNR ambiguities and offsets are continuously crosschecked using TDIR and GOR (lower part of \cref{fig:sensor-fusion}). Both ranging processing and crosschecks rely on a preceding initial data treatment (upper part of \cref{fig:sensor-fusion}), in which the various delays and offsets are compensated for. Ranging processing and crosschecks can be categorized into four parts demonstrated below: PRNR ambiguity, noise reduction, PRNR offset, and SBR ambiguity.\par
The RSF delivers accurate and precise pseudo\-range estimates. To put the RSF into the context of LISA data processing we revisit \cref{fig:tdi-topologies}: in the baseline TDI topology, the pseudorange estimates from the RSF are disentangled into light travel times and differential timer offsets (see \cref{eq-ap:Pseudorange-TCB}). The differential timer offsets are used to synchronize the interferometric measurements, the light travel times serve as delays in TDI. In the alternative topology TDI is executed on the unsynchronized beatnotes. Here the pseudorange estimates from the RSF are directly used as delays in TDI, after they have been combined with on-board delays according to \cref{eq:TDI-delay}.
\subsection{PRNR ambiguity}
As part of the ranging processing, the PRNR needs to be steadily unwrapped: due to the finite PRN code length, the PRNR jumps back to \SI{0}{\kilo\m} when crossing \SI{400}{\kilo\m} and vice versa (see upper plot in \cref{fig:PRNR-Unwrapping}). These jumps are unphysical but easy to identify and to remove. Apart from that, the PRNR ambiguities need to be crosschecked regularly. For that purpose we propose two independent methods below.\par
GOR represents an unambiguous pseudorange observa\-ble. Hence, the combination of PRNR and GOR enables an identification of the PRNR ambiguity integers $a^{\text{prn}}_{ij}$: 
\begin{align}
	&\text{GOR}_{ij}^t(t) - \text{PRNR}_{ij}^{\hat{\tau}_{i}}(\tau) = N^{\text{gor}}_{ij} + a^{\text{prn}}_{ij}(\tau) \cdot \SI{400}{\kilo\m}\nonumber\\
	&\hspace{2cm}+\underbrace{R_{ij}^{t}(t) - R_{ij}^{\hat{\tau}_{i}}(\tau) - O^{\text{prn}}_{ij} - N^{\text{prn}}_{ij}(\tau)}_{\text{negligible}},\\
	&a^{\text{prn}}_{ij}(\tau) = \text{round}\left[\frac{\text{GOR}_{ij}^t(t) - \text{PRNR}_{ij}^{\hat{\tau}_{i}}(\tau)}{\SI{400}{\kilo\m}}\right], \label{eq:PRN-ambiguity-GOR}
\end{align}
\SI{400}{\kilo\m} is the value we assumed for the PRN code length. However, this procedure only succeeds if $\vert N^{\text{gor}}_{ij} \vert$ does not exceed the PRN code's half length, i.e., \SI{200}{\kilo\m}. Other\-wise, a wrong value for the associated PRN ambiguity integer is selected resulting in an estimation error of \SI{400}{\kilo\m} in the corresponding link. Note that $\text{GOR}_{ij}^t(t)$ and $\text{PRNR}_{ij}^{\hat{\tau}_{i}}(\tau)$ are sampled according to different time frames, but this desynchronization is negligible considering the low accuracy required here (see \cref{sec:GOR}).\par
TDIR constitutes an unambiguous pseudorange observable too. Hence, it can be applied as an independent crosscheck of the PRNR ambiguities. We linearly detrend the ISI, RFI, and TMI beatnotes. We then form the first-generation TDI Michelson variables (see \cref{eq:TDI-Michelson-1}) assuming a constant model for the delays. The pseudo\-ranges are actually drifting by \si{10} to \SI{100}{\m \s \tothe{-1}} mainly due to differen\-tial USO frequency offsets (see central plot in \cref{fig:PRNR-Unwrapping}). Therefore, we choose a short integration time (we use \SI{150}{\s}), otherwise the constant delay model is not sufficient. We use the GOR estimates from above as initial delay values in the TDIR estimator. The TDIR estimator then estimates the six pseudoranges as constants, which can be used to resolve the PRNR ambiguity integers:
\begin{align}
	a^{\text{prn}}_{ij}(\tau) = \text{round}\left[\frac{\text{TDIR}_{ij}^{\hat{\tau}_{i}}(\tau) - \text{PRNR}_{ij}^{\hat{\tau}_{i}}(\tau)}{\SI{400}{\kilo\m}}\right]. \label{eq:PRN-ambiguity-TDIR}
\end{align}
It is not necessary to apply second-generation TDI, the first-generation already accomplishes the task (see \cref{Sim-PRN-ambiguity-resolution}).
\subsection{Noise reduction}\label{sec:sensor-fusion-KF-noise}
For the ranging noise reduction in the ranging proces\-sing, we propose to combine PRNR and sideband range rates in a modified version of a linearized Kalman filter (KF). The conventional KF requires all measurements to be sampled according to one universal time grid. However, in LISA each SC involves its own SCET. We circumvent this difficulty by splitting up the system and build one KF per SC. Each KF only processes the measurements taken on its associated SC, so that the individual SCETs serve as time-grids.\par 
The state vector of the KF belonging to SC 1 and its associated linear system model can be expressed as
\begin{align}
	x^{\hat{\tau}_1} &= (R^{\hat{\tau}_1}_{12},\:R^{\hat{\tau}_1}_{13},\:\dot{R}^{\hat{\tau}_1}_{12},\:\dot{R}^{\hat{\tau}_1}_{13},\:\ddot{R}^{\hat{\tau}_1}_{12},\:\ddot{R}^{\hat{\tau}_1}_{13})^\intercal, \\
	x^{\hat{\tau}_1}_{k+1} &= \begin{pmatrix}
	1 & 0 & \Delta t & 0 & \frac{\Delta t^2}{2} & 0\\
	0 & 1 & 0 & \Delta t & 0 & \frac{\Delta t^2}{2}\\
	0 & 0 & 1 & 0 & \Delta t & 0\\
	0 & 0 & 0 & 1 & 0 & \Delta t\\
	0 & 0 & 0 & 0 & 1 & 0\\
	0 & 0 & 0 & 0 & 0 & 1
	\end{pmatrix} \cdot x^{\hat{\tau}_1}_{k} + w^{\hat{\tau}_1}_k,\label{eq:SM}
\end{align}
$k$ being a discrete time index. Eq. \ref{eq:SM} describes the time evolution of the state vector from $k$ to $k+1$. $w^{\hat{\tau}_1}_k$ denotes the process noise vector. In our implementation its covariance matrix $W$ is set
\begin{alignat}{2}
	\text{E} \left[ w_k \cdot w_l^{\text{T}} \right] &= \delta_{k,\:l} \: W, \label{eq:Process-Noise-White}\\
	W &= \text{diag}\Big{(}&&0,\:0,\:0,\:0,\nonumber\\
	& &&10^{-15}\si{\s\tothe{-1}},\:10^{-15}\si{\s\tothe{-1}}\Big{)}^{2},\label{eq:Process-Noise-Cov-Mat}
\end{alignat}
$\delta_{k,\:l}$ denotes the Kronecker delta. The measurement vector and the associated observation model are given by
\begin{align}
	y^{\hat{\tau}_1} &= (\text{PRNR}^{\hat{\tau}_1}_{12},\: 	\text{PRNR}^{\hat{\tau}_1}_{13},\: \dot{\text{SBR}}^{\hat{\tau}_1}_{12},\: \dot{\text{SBR}}^{\hat{\tau}_1}_{13})^\intercal, \\
	y^{\hat{\tau}_1}_k &= \begin{pmatrix}
	1 & 0 & 0 & 0 & 0 & 0\\
	0 & 1 & 0 & 0 & 0 & 0\\
	0 & 0 & \SI{2.401}{\giga\hertz} & 0 & 0 & 0\\
	0 & 0 & 0 & \SI{2.400}{\giga\hertz} & 0 & 0
	\end{pmatrix} \cdot x^{\hat{\tau}_1}_k + v^{\hat{\tau}_1}_k.\label{eq:OM}
\end{align}
Eq. \ref{eq:OM} relates the measurement vector to the state vector. $v^{\hat{\tau}_1}_k$ denotes the measurement noise vector. In our implementation its covariance matrix $V$ is set
\begin{alignat}{2}
	\text{E} \left[ v_k \cdot v_l^{\text{T}} \right] &= \delta_{k,\:l} \: V,\\
	V&=\text{diag}\Big{(}&&3\cdot 10^{-9}\si{\s}, \:3\cdot 10^{-9}\si{\s},\nonumber\\
	&\: &&5.2\cdot 10^{-13},\:\:5.2\cdot 10^{-13}\Big{)}^2.\label{eq:Measurement-Noise-Cov-Mat}
\end{alignat}
The diagonal entries denote the variances of the respective measurements. We assume the measurements to be uncorrelated, so that the off-diagonal terms are zero. The KFs for SC 2 and SC 3 are defined accordingly. Hence, we remove the ranging noise and obtain estimates for all the six pseudo\-ranges and their time derivatives.\par 
These pseudorange estimates are dominated by the right-handed modulation noise, which is one order of magnitude higher than the left-handed one. As pointed out in \cite{Hartwig:TDIwoSync}, the right-handed modulation noise can be subtracted (see \cref{appendix:RH-modulation-noise}): we combine the RFI measurements to form the $\Delta M_{i}$, which are measurements of the right-handed modulation noise on SC $i$ (see \cref{eq:reference-sbs}). For right-handed MOSAs, the local right-handed modulation noise enters the sideband range rates and we just need to subtract the local $\Delta M_i$ (see \cref{eq:Reduced-RH-Modnoise-No-Delay}). For left-handed MOSAs the Doppler-delayed right-handed modu\-lation noise from the distant SC appears in the sideband range rates. Here we need to apply the Kalman filter estimates for the pseudoranges and their time derivatives to form the Doppler-delayed distant $\Delta M_i$, which then can be subtracted (see \cref{eq:Reduced-RH-Modnoise-Delay}). We then process the three KFs again, this time with the corrected sideband range rates. Now they are limited by left-handed modulation noise, so that the respective noise levels are lower. Therefore, we need to adjust the measurement noise covariance matrix for the second run of the KFs:
\begin{alignat}{2}
	V_{\text{ cor}}&=\text{diag}\Big{(}&&3\cdot 10^{-9}\si{\s}, \:3\cdot 10^{-9}\si{\s},\nonumber\\
	&\: &&7.4\cdot 10^{-14},\:\:7.4\cdot 10^{-14}\Big{)}^2.\label{eq:Measurement-Noise-Cov-Mat-Red}
\end{alignat}
In this way we obtain estimates for the pseudoranges and their time derivatives, which are limited by the left-handed modulation noise.
\subsection{PRNR offset}\label{sec:sensor-fusion-tdir-offset}
The PRNR offset is calibrated on ground before mission start. During operation, it is constructed with the help of SC monitors and subtracted in the initial data treatment.\par 
TDIR can be used as a crosscheck for residual PRNR offsets, as it is sensitive to offsets in the delays. To obtain optimal performance we choose the second-generation TDI Michelson variables (see \cref{eq:TDI-Michelson-2}) to be ultimately limited by secon\-dary noises. In-band clock noise is sufficiently suppressed, since we operate on beatnotes in total frequency and make use of the in-band ranging information provided by the preceding noise reduction step. Accordingly, the offset delay model is parameterized by
\begin{align}\label{eq:TDIR-Offset-Estimator-Delay-Model}
	d_{ij}^{\hat{\tau}_i}(\tau) = \hat{R}_{ij}^{\hat{\tau}_i}(\tau) - O_{ij},
\end{align}
$\hat{R}_{ij}^{\hat{\tau}_{i}}$ denote the pseudorange estimates after noise reduction, $O_{ij}$ are the 6 offset parameters. As discussed in \cref{sec:TDIR}, computing TDI in total frequency units gene\-rally results in a variable with residual trends. Those trends need to be removed prior to calculation of the TDIR integral to be sensitive to residual laser noise in band. This is achieved by an appropriate band-pass filter with a pass-band from \SIrange{0.1}{1}{\Hz}. The TDIR integral then reads
\begin{align}\label{eq:TDIR-integral-freq}
	\hat{O}_{ij} = \argmin_{O_{ij}}\int_{0}^{T}\!\tilde X^2(t) + \tilde Y^2(t) + \tilde Z^2(t)\,\mathrm{d}t
\end{align}
where tilde indicates the filtered quantities. $\hat{O}_{ij}$ are the estimated offset parameters.
\subsection{SBR ambiguity}
Phase anchor points in combination with the pseudo\-range estimates after noise reduction enable the SBR ambiguity resolution (see \cref{eq:Sideband-Ranging}):
\begin{align}\label{eq:SBR-Ambiguity-Resolution}
	a^{\text{sb}}_{ij}(\tau) = \text{round}\left[\nu^{\text{m}}_{ji}\:\hat{R}_{ij}^{\hat{\tau}_{i}}(\tau) - \text{SBR}_{ij}^{\hat{\tau}_{i}}(\tau)\right],
\end{align}
$\text{SBR}_{ij}^{\hat{\tau}_{i}}$ are the phase anchor points, $\hat{R}_{ij}^{\hat{\tau}_{i}}$ the pseudorange estimates of the core ranging processing pipeline. Thus, we obtain estimates of the SBR ambiguity integers $a^{\text{sb}}_{ij}$. These can be used to compute unambiguous SBR pseudo\-range estimates associated to the phase anchor points, which serve as initial values for the integration of the sideband range rates (\cref{eq:dotSBR}). This procedure is successful if the $\hat{R}_{ij}^{\hat{\tau}_{i}}$ are more accurate than \SI{6.25}{\centi \m} (half the SBR ambiguity). Then, SBR constitutes a very accurate pseudo\-range observable, as not only its precision but also its accuracy are limited by the modulation noise, in contrast to the $\hat{R}_{ij}^{\hat{\tau}_{i}}$, whose accuracy is ultimately limited by the ranging noise. We assume here that the SBR offset (\cref{eq:SBR-offset}) is corrected in the initial data treatment.

\section{Results}\label{sec:results}
\begin{figure}
	\begin{center}
		\includegraphics[width=0.5\textwidth]{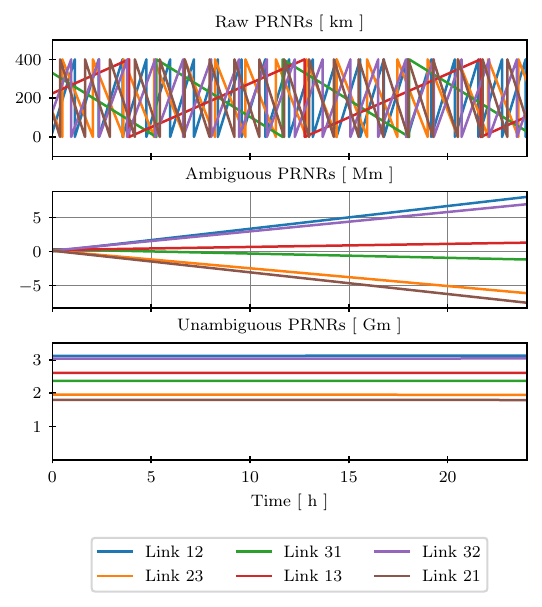}
	\end{center}
	\caption{Upper plot: raw PRNR. The ambiguity jumps at \SI{0}{\kilo\m} and \SI{400}{\kilo\m} can be seen. Central plot: ambiguous PRNR, the jumps have been removed but the PRNR ambiguities have not been resolved yet. The large slopes are mainly due to USO frequency offsets. Lower plot: unambiguous PRNR. The large differences between the links are caused by differential SCET offsets.}\label{fig:PRNR-Unwrapping}
\end{figure}
In this section, we demonstrate the performance of our implementation of the core ranging processing and the crosschecks as proposed in \cref{sec:sensor-fusion} (central and lower part of \cref{fig:sensor-fusion}). We did not implement the initial data treatment. Instead we assume that the PRNR and sideband timestamping delays are compensated beforehand. We further consider offset-free PRNR and apply TDIR as a crosscheck for residual offsets.\par
\begin{figure*}
	\begin{center}
		\includegraphics[width=1\textwidth]{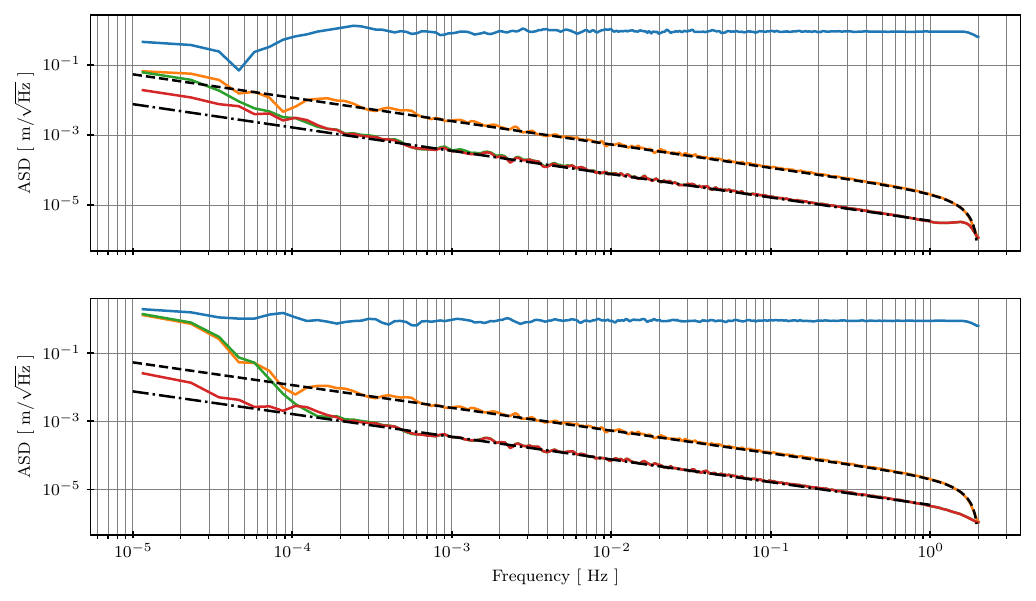}
	\end{center}
	\caption{ASDs of the residual pseudorange estimates for link 12 (upper plot) and link 21 (lower plot). In blue, residual PRNR. In orange, residual pseudorange estimates after ranging noise reduction. In green, residual pseudorange estimates after subtraction of right-handed modulation noise. In red, residual SBR. Dashed black lines, right-handed modulation noise model. Dash-dotted black lines, left-handed modulation noise model.}\label{fig:residual-noise-asds}
\end{figure*}
We use telemetry data simulated by LISA Instrument \cite{bayle_jean_baptiste_2022_7071251} and LISANode \cite{bayle_jean_baptiste_2022_6461078} based on orbits provided by ESA \cite{MartensOD, bayle_jean_baptiste_2022_7700361}. We simulate phase anchor points for the SBR (see \cref{eq:Sideband-Ranging}). The SCET deviations from the respective proper times are modeled as
\begin{align}
	\delta \hat{\tau}_{i}(\tau) = \delta \hat{\tau}_{i,\:0} + y_i \: \tau + \frac{\dot{y}_i}{2} \: \tau^2 + \frac{\ddot{y}_i}{3} \: \tau^3 + \int_{\tau_0}^{\tau} \text{d} \tilde{\tau} \: y_{i}^{\epsilon}(\tilde{\tau}),
\end{align}
the $\delta \hat{\tau}_{i,\:0}$ denote the initial SCET deviations set to \SI{1}{\s}, \SI{-1.2}{\s}, and \SI{0.6}{\s} for SC 1, 2, and 3, respectively. The $y_i$ model the PMC frequency offsets corresponding to linear clock drifts. They are set to $10^{-7}$, $-2 \times 10^{-7}$, and $0.6 \times 10^{-7}$ for SC 1, 2, and 3, respectively. $\dot{y}_i \sim 10^{-14}\:\si{\s\tothe{-1}}$ and $\ddot{y}_i \sim 10^{-23}\:\si{\s\tothe{-2}}$ are constants modeling the linear and quadratic PMC frequency drifts. The $y_{i}^{\epsilon}$ denote the stochastic clock noise in fractional frequency deviations, the associated ASD is given by
\begin{align}
	\sqrt{S_{y^{\epsilon}}(f)} = 6.32 \times 10^{-14} \si{\hertz\tothe{-0.5}} \left(\frac{f}{\si{\hertz}}\right)^{-0.5}.
\end{align}
We simulate laser frequency noise with an ASD of
\begin{align}
	\sqrt{S_{\dot{N}^{p}}(f)} = \SI{30}{\hertz \hertz\tothe{-0.5}},\label{eq:ASD-laser-noise}
\end{align}
and ranging and modulation noise as specified in the sections \ref{sec:PRN-Ranging} and \ref{sec:SBR}. Furthermore, we consider test-mass acceleration noise
\begin{align}
	\sqrt{S_{N^{\delta}}(f)} &= 4.8 \times 10^{-15}\si{\m\s\tothe{-2}\hertz\tothe{-0.5}} \sqrt{1 + \left(\frac{\SI{0.4}{\milli\hertz}}{f}\right)^2}
\end{align}
and readout noise 
\begin{align}
	\sqrt{S_{N^{\text{ro}}}(f)} = A \: \sqrt{1 + \left(\frac{\SI{2}{\milli\hertz}}{f}\right)^4},
\end{align}
where $A=6.35 \times 10^{-12} \si{\m\hertz\tothe{-0.5}}$ for the ISI carrier and $A=1.25 \times 10^{-11} \si{\m\hertz\tothe{-0.5}}$ for the ISI sideband beatnotes. For the readout noise we set a saturation frequency of $f_{\text{sat}}=\SI{0.1}{\milli\hertz}$, below which we whiten. The orbit determinations are simulated by LISA Ground Tracking with the noise levels specified in \cref{sec:GOR}.
\subsection{Ranging processing}
Here we demonstrate the performance of our implementation of the core ranging processing for one day of telemetry data simulated by LISA Instrument \cite{bayle_jean_baptiste_2022_7071251}. The first ranging processing step covers the PRNR unwrapping (see \cref{fig:PRNR-Unwrapping}). The upper plot shows the raw PRNR, which jumps back to \SI{0}{\kilo\m} when crossing \SI{400}{\kilo \m} and vice versa. These jumps are easy to identify and to remove. In our implementation we remove all PRNR jumps bigger than \SI{200}{\kilo\m}. The central plot shows the unwrapped but ambiguous PRNR. Here we see PRNR drifts of the order of \si{10} to \SI{100}{\m \s\tothe{-1}}, which are mainly due to differential USO frequency offsets. Inserting the PRNR ambiguity integers obtained from GOR and TDIR yields the unambiguous PRNR shown in the lower plot.\par
In the second step, we use the Kalman filter presented in \cref{sec:sensor-fusion} to reduce the ranging noise. Subsequently, we subtract the right-handed modulation noise applying the $\Delta M$ measurements constructed from the RFI beatnotes (see \cref{appendix:RH-modulation-noise}). After noise reduction, we resolve the SBR ambiguities combining the estimated pseudo\-ranges with the simulated SBR phase anchor points (see \cref{eq:SBR-Ambiguity-Resolution}). We then integrate the sideband range rates, to obtain unambiguous SBR.\par
\begin{figure}
	\begin{center}
		\includegraphics[width=0.5\textwidth]{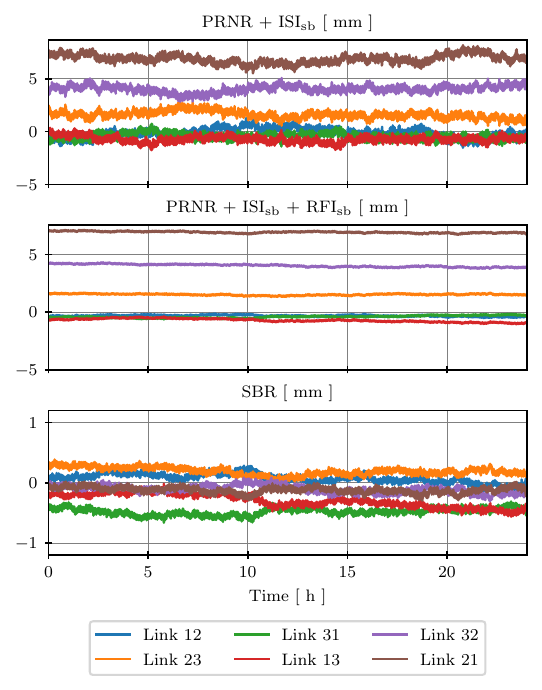}
	\end{center}
	\caption{Upper plot: residual pseudorange estimates after ranging noise reduction. Second plot: residual pseudorange estimates after subtraction of right-handed modulation noise. Third plot: residual SBR estimates.}\label{fig:Residual-PR-Estimates}
\end{figure}

In \cref{fig:residual-noise-asds}, we plot the ASDs of the residual pseudo\-range estimates (deviations of the estimates from the true pseudorange values in the simulation) for link 12 (upper plot) and link 21 (lower plot). Blue lines show the ASDs of the residual PRNR, which are essentially the ASDs of the white ranging noises. The residual pseudorange estimates after ranging noise reduction are plotted in orange. They are obtained by combining the PRNR with the sideband range rates. Therefore, they are limited by right-handed modulation noise (dashed black line). In green, we plot the residual pseudorange estimates after subtraction of right-handed modulation noise with the RFI beatnotes. Now the estimates are limited by left-handed modulation noise (dash-dotted black line). The residual SBR are drawn red, they are limited by left-handed modulation noise as well, but involve a smaller offset, since the SBR phase anchor points are more accurate than PRNR after ranging noise reduction (see \cref{fig:Residual-PR-Estimates}). In the case of left-handed MOSAs (see link 12) the RFI beatnotes need to be time shifted to form the delayed $\Delta M$ measurements. We apply the time shifting method of PyTDI \cite{staab_martin_2023_7704609}, which consists in a Lagrange interpolation (we use order 5). The interpolation introduces noise in the high frequency band (see the bump at \SI{2}{\hertz} in the upper plot) but this is out of band.\par
Fig. \ref{fig:Residual-PR-Estimates} shows the different residual pseudorange estimates as time series. The upper plot shows the 6 residu\-al pseudorange estimates after ranging noise reduction, the second plot after subtraction of right-handed modulation noise. The third plot shows the SBR residuals. The subtraction of right-handed modulation noise reduces the noise floor, but it does not increase the accuracy of the pseudorange estimates. The accuracy can be increased by one order of magnitude through the resolution of the SBR ambiguities. After ambiguity resolution, SBR constitutes pseudorange estimates with sub-\si{\milli\m} accuracy.
\begin{figure*}
	\begin{center}
		\includegraphics[width=1\textwidth]{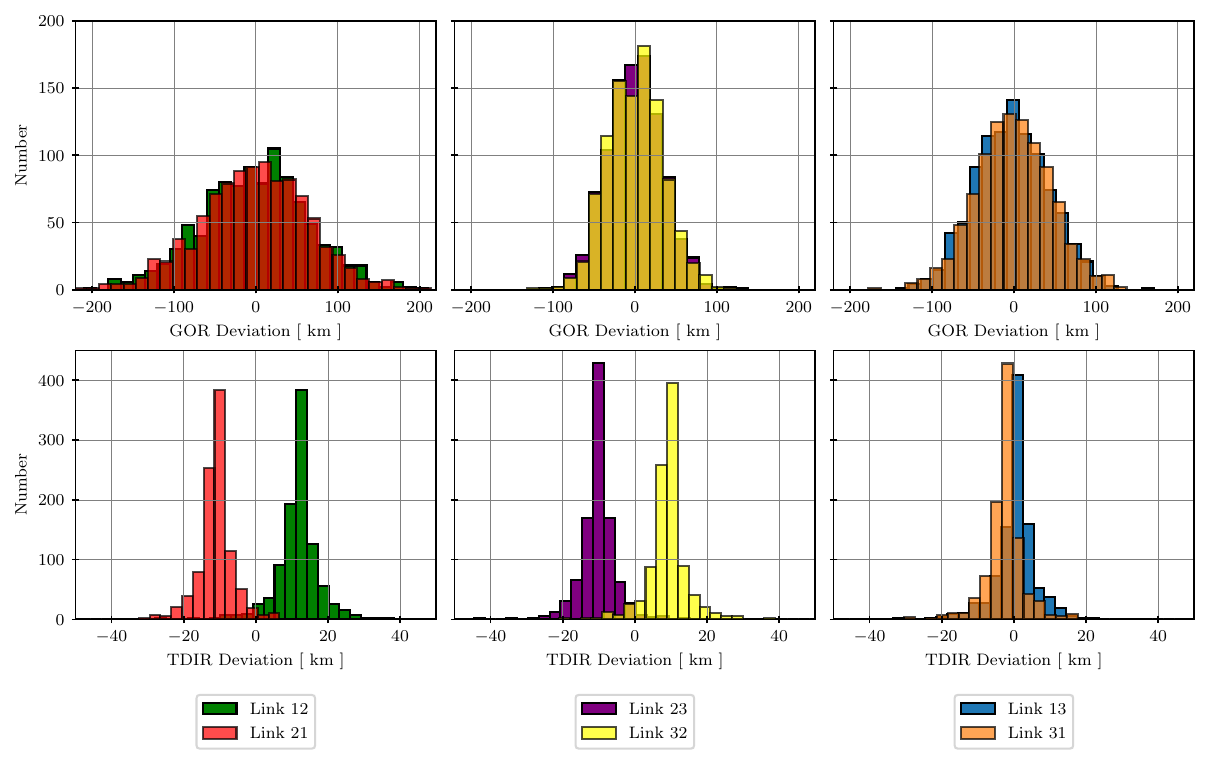}
	\end{center}
	\caption{PRNR ambiguity resolution via GOR (upper plots) and TDIR (lower plots). The histogram plots show the residual GOR and TDIR pseudorange estimates (deviations from true pseudorange) for the different links.}\label{Sim-PRN-ambiguity-resolution}
\end{figure*}
\subsection{Crosschecks}
Here we demonstrate the performance of our implementation of the crosschecks for PRNR ambiguity and PRNR offset.\par
The PRNR ambiguities can be resolved using either GOR (see \cref{eq:PRN-ambiguity-GOR}) or TDIR (see \cref{eq:PRN-ambiguity-TDIR}).
To evaluate the performance of both methods, we simulate 1000 short (\SI{150}{\s}) telemetry datasets with LISA Instrument \cite{bayle_jean_baptiste_2022_7071251}, and one set of ODs and MOC-TCs for each of them. We compute the GOR and TDIR pseudorange estimates for each of the 1000 datasets. Fig. \ref{Sim-PRN-ambiguity-resolution} shows the GOR residu\-als (first row) and the TDIR residuals (second row) in \si{\kilo \m} as histogram plots. We see that the GOR accuracy depends on the arm, because we obtain more accurate ODs for arms oriented in line of sight direction than for those oriented cross-track. The PRNR ambiguity resolution via GOR is successful for GOR deviations smaller than \SI{200}{\kilo \m}.
\begin{figure*}
	\begin{center}
		\includegraphics[width=1\textwidth]{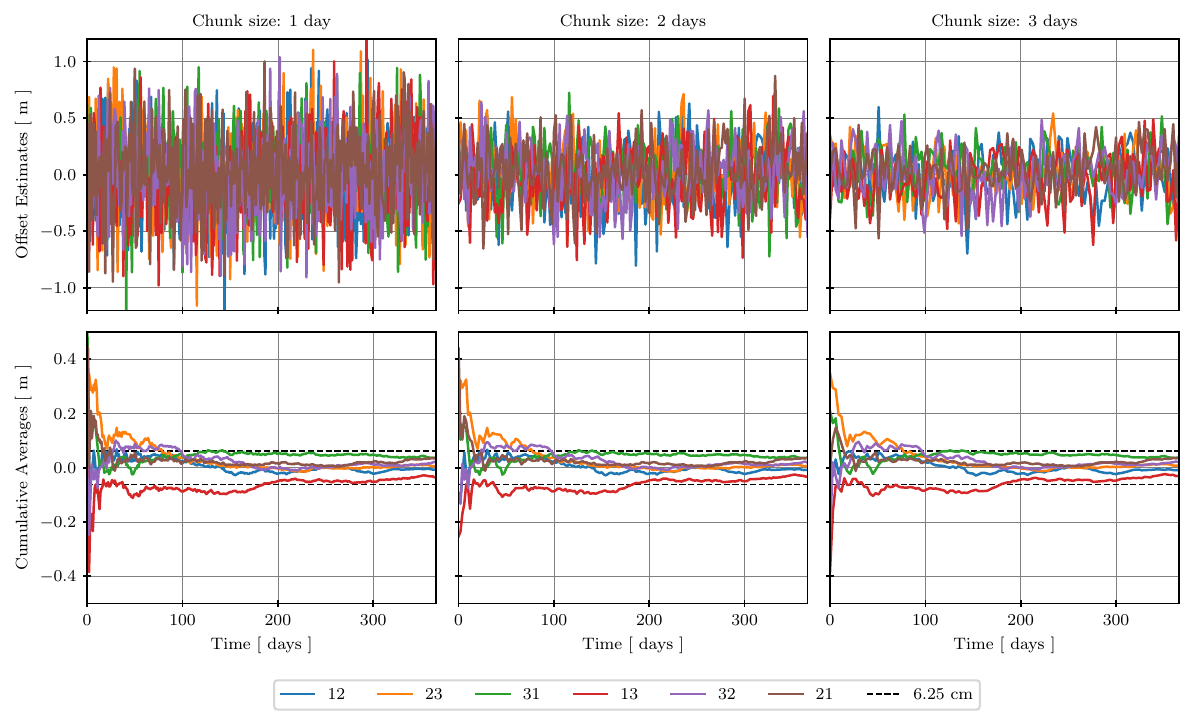}
	\end{center}
	\caption{We simulate one year of telemetry data with PRNR offsets in the order of \SI{100}{\m}. We divide this dataset into 1 day chunks (left plots), 2 day chunks (central plots), and 3 day chunks (right plots). We then we apply TDIR to each of these chunks in order to estimate the PRNR offsets. Upper plots: Residual offset estimates in \si{\m} for the different chunk sizes. Lower plots: residual offset estimates after cumulative averaging in \si{\m} for the different chunk sizes. Dashed-black lines: half the SBR ambiguity.}\label{fig:TDIR-Offset-Estimation}
\end{figure*}
In the case of the links 23, 31, 13, and 32 all PRNR ambiguity resolutions via GOR are successful. For each of the links 12 and 21, 2 out of the 1000 PRNR ambiguity resolutions fail. The GOR estimates are passed as initial values to TDIR, which then reduces the uncertainty by almost one order of magnitude (lower plot in \cref{Sim-PRN-ambiguity-resolution}), such that eventually all PRNR ambiguity resolutions are successful. The direction dependent offsets we observe in the TDIR estimates are due to the fact that we apply constant delays in the model used in the minimization of \cref{eq:TDIR-integral-freq}. In reality, these delays drift with time mainly due to the differential USO frequency offsets, and these drifts are opposite for counterpropagating links (central plot in \cref{fig:PRNR-Unwrapping}). The offsets are not a problem, since the corresponding estimates are one order of magnitude more accurate than what we need.\par
TDIR can also be applied to estimate the PRNR offsets. Hence, it constitutes a cross-check of the on-ground PRNR offset calibration. We simulate one year of telemetry data using LISANode \cite{bayle_jean_baptiste_2022_6461078}. We set the PRNR offsets to \SI{160.3}{\m}, \SI{-210.2}{\m}, \SI{137.3}{\m}, \SI{-250.3}{\m}, \SI{-188.8}{\m}, and \SI{105.1}{\m} for the links 12, 23, 31, 13, 32, and 21, respectively. We divide the dataset into 1 day chunks (left plots in \cref{fig:TDIR-Offset-Estimation}), 2 day chunks (central plots in \cref{fig:TDIR-Offset-Estimation}), and 3 day chunks (right plots in \cref{fig:TDIR-Offset-Estimation}). In each partition we apply the TDIR estimator presented in \cref{sec:sensor-fusion-tdir-offset} to each chunk in order to estimate the PRNR offsets. This computation was parallelized and executed on the ATLAS cluster at the AEI Hannover. In the upper part of \cref{fig:TDIR-Offset-Estimation} we show the offset estimation residuals for the three chunk sizes. The offset estimation accuracy increases with the chunk size in agreement with the order of magnitude estimate through \cref{eq:TDIR-accuracy}. In the lower part of \cref{fig:TDIR-Offset-Estimation} we plot the residual cumulative averages of the PRNR offset estimates for the different chunk sizes. Here, it can be seen that the TDIR estimator performs similarly for the different chunk sizes. With the 3 day chunk size we can estimate all PRNR offsets with an accuracy of better than \SI{20}{\centi\m} after 10 days. The dashed-black lines indicate \SI{6.25}{\centi\m} (half the SBR ambiguity). This is the required PRNR offset estimation accuracy for a successful SBR ambiguity resolution. All offset estimation residuals are below these \SI{6.25}{\centi\m} after 179 days.

\section{Conclusion}\label{sec:conclusion}
The on-board ranging system PRNR requires three treatments due to its ambiguity, offset, and noise. In this article we propose a ranging sensor fusion (RSF), which uses three further observables to solve these issues in order to obtain accurate and precise pseudorange estimates. We show that both GOR and TDIR enable the resolution of the PRNR ambiguity. We investigate how on-board delays affect the pseudorange observables, and propose an initial data treatment to compensate their effects (offsets, timestamping delays) based on an on-ground cali\-bration of the on-board delays. We implement TDIR as a crosscheck for the PRNR offset cali\-bration. We use a Kalman filter to remove the white ranging noise combi\-ning PRNR and sideband range rates and apply the RFI beatnotes to subtract the right-handed modulation noise. This leads to pseudorange estimates at sub-\si{\centi\m} accuracy. We show that in combination with phase anchor points they allow to resolve the SBR ambiguities resulting in pseudorange estimates at sub-\si{\milli\m} accuracy. Apart from that, we identify the delays that are to be applied in TDI in consideration of on-board delays. These are the pseudoranges in combination with on-board delays on both the emitter and the receiver side (see \cref{eq:TDI-delay}).\par
TDI requires a ranging accuracy of about \SI{10}{\m} to suppress the laser frequency noise below the secondary noise levels \cite{TintoTDIAccuracy}. While the main building blocks of the here presented RSF are necessary, the reached sub-\si{\milli\m} ranging accuracy does not translate into a higher sensitivi\-ty for gravitational waves in the final output. Nevertheless, a better ranging accuracy is beneficial: the secondary noise levels might decrease during the decade until launch; in the case that the cavities perform below expectations, i.e., the encountered laser frequency noise is higher than expected, TDI needs a better ranging accuracy; also in the context of next generation space-based gravitational wave missions a better ranging accuracy might be advantageous.\par
From the economic perspective, it could be argued whether the on-board ranging system could be dropped in favor of TDIR. Hence, it must be emphasized that PRNR is the most reliable LISA pseudorange observa\-ble. After ambiguity resolution and offset correction PRNR delivers a sub-\si{\m} ranging accuracy directly with the first sample. The TDIR estimator needs more than \SI{1000}{s} of telemetry data to reach a similar accuracy (see \cref{eq:TDIR-accuracy}). Hence, considering the possibility of data gaps, it is convenient to have PRNR as a direct pseudorange observa\-ble. Furthermore, PRNR is more robust than TDIR, which relies on the science data laser frequency noise being its signal. Not only the secondary noise sources but also e.g., gravitational wave signals, are noise to TDIR and degrade its performance. Finally, the PRN modulation has since long been a part of the LISA baseline architecture, as it serves another indisputable function: data transfer between the SC at 45 to 60 kbits \si{\s\tothe{-1}}. This requires some kind of digital phase modulation anyway. Implementing the ranging function on top imposes only uncritical extra constraints on that modulation, i.e., synchronization of the digital codes to the SCET.\par
While we carefully investigate the effects of on-board delays in the ISI, we neglect them in the RFI and in the TMI. However, any differential delay between the laser frequency noise terms appearing in the ISI and in the RFI will cause residual laser frequency noise. Therefore, a follow-up study should investigate the effects of on-board delays for the full LISA interferometeric system.\par
In the PRNR offset estimation via TDIR we assume these offsets to be constant. In reality, however, they are expected to be slowly time-varying. In a follow-up study, the PRNR offset estimation via TDIR should be extended to linearly time varying PRNR offsets. The delay model for the TDIR estimator would then become:
\begin{align}
	d_{ij}^{\hat{\tau}_i}(\tau) = \hat{R}_{ij}^{\hat{\tau}_i}(\tau) - (O^0_{ij} + O^1_{ij} \cdot \tau).
\end{align}
The TDIR estimator would now have to fit the 6 constants $O^0_{ij}$ and the 6 drifts $O^1_{ij}$. Tone-assisted TDIR \cite{francis2015tone} could be included in this study to reach a better accuracy.\par
Both, the Kalman filter algorithm presented in \cref{sec:sensor-fusion-KF-noise} and the TDIR estimator presented in \cref{sec:sensor-fusion-tdir-offset} are pseudorange estimators. A follow-up study could investigate whether their combination in a single estimator leads to better pseudorange estimates \cite{Baghi2023-PCA}.\par
Furthermore, it would be interesting to include time-varying on-board delays and the associated SC monitors into the simulation. This would enable an inspection of the feasibility of the initial data treatment as proposed in \cref{sec:sensor-fusion}.\par 
Finally, the RSF could be included into the different INReP topologies. Apart from that, the algorithms could be applied to real data as, e.g., produced by the hexagon experiment \cite{Yamamoto:HexagonSync}, \cite{Schwarze:Hexagon}.

\section*{Acknowledgements}
J. N. R. acknowledges the funding by the Deutsche Forschungsgemeinschaft (DFG, German Research Foundation) under Germany's Excellence Strategy within the Cluster of Excellence PhoenixD (EXC 2122, Project ID 390833453). Furthermore, he acknowledges the support by the IMPRS on Gravitational Wave-Astronomy at the Max Planck Institute for Gravitational Physics in Hannover, Germany. This work is also supported by the Max-Planck-Society within the LEGACY (``Low-Frequency Gravitational-Wave Astronomy in Space'') collaboration (M.IF.A.QOP18098). O. H. and A. H. acknowledge support from the Programme National GRAM of CNRS/INSU with INP and IN2P3 co-funded by CNES and from the Centre National d'Études Spatiales (CNES). The authors thank Miles Clark, Pascal Grafe, Waldemar Martens, and Peter Wolf for useful discussions. The study on PRNR offset estimation via TDIR was performed on the ATLAS cluster at AEI Hannover. The authors thank Carsten Aulbert and Henning Fehrmann for their support. 

\appendix
\section{Pseudoranges in TCB}\label{appendix:pr_TCB}
The pseudorange can be expressed in TCB by writing the SCETs of receiving and emitting SC as functions of TCB evaluated at the events of reception and emission, respectively:
\begin{align}
	R_{ij}^t(t_\text{rec})= \hat{\tau}_{i}^{t}(t_{\text{rec}}) - \hat{\tau}_{j}^{t}(t_{\text{emit}}),
\end{align}
$\hat{\tau}_{i}^{t}$ denotes the SCET of SC $i$ expressed as a function of TCB. The TCB of emission can be expressed as the difference between the TCB of reception and the light travel time from SC $j$ to SC $i$, denoted by $d_{ij}^{t}$:
\begin{align}\label{eq-ap:PR-temp}
	R_{ij}^t(t_\text{rec})=\hat{\tau}_{i}^{t}(t_{\text{rec}}) - \hat{\tau}_{j}^{t}\left(t_{\text{rec}} - d_{ij}^{t}(t_{\text{rec}})\right),
\end{align}
in the following we drop the subscript, hence $t$ refers to the TCB of reception. The SCET can be expressed in terms of the SCET deviation from TCB
\begin{align}
	\hat{\tau}_{i}^t(t) &= t +  \delta \hat{\tau}_{i}^t(t),
\end{align}
which  allows us to write \cref{eq-ap:PR-temp} as
\begin{align}
	R_{ij}^t(t)&= \delta\hat{\tau}_{i}^{t}(t) + d_{ij}^{t}(t) - \delta \hat{\tau}_{j}^{t}\left(t - d_{ij}^{t}(t)\right).
\end{align}
Expanding the SCET deviation of the emitting SC from TCB around the reception TCB yields:         
\begin{align}\label{eq-ap:Pseudorange-TCB}
	R_{ij}^t(t)&= \delta\hat{\tau}_{ij}^{t}(t) + \left(1 + \delta \dot{\hat{\tau}}_{j}^{t}(t)\right) \cdot  d_{ij}^{t}(t),\\
	\delta\hat{\tau}_{ij}^{t}(t):&= \delta\hat{\tau}_{i}^{t}(t) - \delta \hat{\tau}_{j}^{t}(t).
\end{align}
Hence, in a global time frame like TCB, the pseudorange can be expressed in terms of the light travel time $d_{ij}^{t}$ and the differential SCET offset $\delta\hat{\tau}_{ij}^{t}$.

\section{Subtraction of right-handed modulation noise}\label{appendix:RH-modulation-noise}
Following the notation in \cite{Hartwig:TDIwoSync}, we express the RFI carrier and sideband beatnotes in frequency:
\begin{align}
	\text{RFI}_{ij}^{\hat{\tau}_i}(\tau) =&\: \nu^{\hat{\tau}_i}_{ik}(\tau) - \nu^{\hat{\tau}_i}_{ij}(\tau),\label{eq:RFI_c}\\
	\text{RFI}_{\text{sb},\:ij}^{\hat{\tau}_i}(\tau)
	=&\:\nu^{\hat{\tau}_i}_{\text{sb},\:ik}(\tau)- \nu^{\hat{\tau}_i}_{\text{sb},\:ij}(\tau),\\
	\nu^{\hat{\tau}_i}_{\text{sb},\:ij}(\tau)=&\:\nu^{\hat{\tau}_i}_{ij}(\tau)+\nu_{ij}^{\text{m}}\cdot(1+M_{ij}^{\hat{\tau}_i}).
\end{align}
We combine these beatnotes to form measurements of the right-handed modulation noise:
\begin{align}
	\Delta M_{i}^{\hat{\tau}_{i}} \vcentcolon =&\: \frac{\text{RFI}_{ij}^{\hat{\tau}_i} - \text{RFI}_{\text{sb},\:ij}^{\hat{\tau}_i} + \SI{1}{\mega\hertz}}{2}\nonumber\\-&\:\frac{\text{RFI}_{ik}^{\hat{\tau}_i} - \text{RFI}_{\text{sb},\:ik}^{\hat{\tau}_i} - \SI{1}{\mega\hertz}}{2},\nonumber\\
	=&\:\nu^{\text{m}}_{ij}\cdot M^{\hat{\tau}_i}_{ij} - \nu^{\text{m}}_{ik} \cdot M^{\hat{\tau}_i}_{ik},\label{eq:reference-sbs}
\end{align}
$i,\:j,$ and $k$ being a cyclic permutation of 1, 2, and 3. We can now subtract the $\Delta M_{i}^{\hat{\tau}_{i}}$ measurements from the sideband range rates (\cref{eq:dotSBR}) to reduce the right-handed modulation noise. After that, we are limited by the one order of magnitude lower left-handed modulation noise:
\begin{subequations}
	\begin{align}
		&\: \dot{\text{SBR}}_{\text{cor},\:ij}^{\hat{\tau}_{i}} = \dot{\text{SBR}}_{ij}^{\hat{\tau}_{i}} - \dot{\textbf{D}}^{\hat{\tau_i}}_{ij} \cdot \Delta M_{j}^{\hat{\tau}_{j}}(\tau),\nonumber\\
		=&\: \nu^{\text{m}}_{ji} \cdot \dot{R}_{ij}^{\hat{\tau}_{i}}
		+\nu^{\text{m}}_{ij} \left( M^{\hat{\tau}_i}_{ij}-\dot{\textbf{D}}^{\hat{\tau_i}}_{ij}\cdot M^{\hat{\tau}_j}_{jk}(\tau )\right),\label{eq:Reduced-RH-Modnoise-Delay}\\
		&\: \dot{\text{SBR}}_{\text{cor},\:ik}^{\hat{\tau}_{i}} = \dot{\text{SBR}}_{ik}^{\hat{\tau}_{i}}(\tau) + \Delta M_{i}^{\hat{\tau}_{i}}(\tau),\nonumber\\
		=&\: \nu^{\text{m}}_{ki} \cdot \dot{R}_{ik}^{\hat{\tau}_{i}}
		+\nu^{\text{m}}_{ki} \left( M^{\hat{\tau}_i}_{ij}+\dot{\textbf{D}}^{\hat{\tau_i}}_{ik}\: M^{\hat{\tau}_k}_{ki}(\tau)\right),\label{eq:Reduced-RH-Modnoise-No-Delay}
	\end{align}
\end{subequations}
$i,\:j,$ and $k$ being a cyclic permutation of 1, 2, and 3.

\section{Solar wind dispersion}\label{appendix:Solar-Wind}
The average solar wind particle density at the LISA orbit is about \SI{10}{\centi\m\tothe{-3}}. Hence, at the scales of optical wavelengths the solar wind plasma can be treated as a free electron gas with the plasma frequency \cite{Smetana:Solar-Wind}
\begin{align}
	\nu_p^2 = \frac{n_e\:e^2}{4\pi^2 \: \epsilon_0\:m_e} \approx 8 \times 10^{8}\: \si{\s\tothe{-2}},
\end{align}
$n_e$ denotes the electron density, $e$ the elementary charge, $m_e$ the electron mass, and $\epsilon_0$ the vacuum permittivi\-ty. Contributions from protons and ions can be neglected as the plasma frequency is inversely proportional to the mass. We describe the refractive index of the solar wind plasma by the Appleton equation. Neglecting collisions and magnetic fields it denotes
\begin{align}
	n(\nu) = \sqrt{1 - \left(\frac{\nu_p}{\nu}\right)^2}.
\end{align}
In a dispersive medium we need to distinguish between phase and group velocity. The phase velocity is given by
\begin{align}
	v_\text{p}(\nu)& = \frac{c}{n(\nu)} = \frac{c}{\sqrt{1 - \left(\frac{\nu_p}{\nu}\right)^2}} \approx c \cdot \left(1 + \frac{1}{2} \frac{\nu_p^2}{\nu^2}\right),
\end{align}
where we applied the expansion for $\nu \gg \nu_p$, as we consider optical frequencies. The product of group and phase velocity yields $c^2$. Consequently, the group velocity is
\begin{align}
	v_\text{g}(\nu)& = c \cdot n(\nu) = c \cdot \sqrt{1 - \left(\frac{\nu_p}{\nu}\right)^2} \approx c \cdot \left(1 - \frac{1}{2} \frac{\nu_p^2}{\nu^2}\right).
\end{align}
The group and the phase delay for the interspacecraft signal propagation in LISA can now be expressed as
\begin{align}
	\Delta \tau_{\text{g}}(\nu) &= L \left(\frac{1}{c \cdot \sqrt{1 - \left(\frac{\nu_p}{\nu}\right)^2}} - \frac{1}{c} \right) \approx \frac{L\:\nu_p^2}{2\:c} \cdot \frac{1}{\nu^2},\\
	\Delta \tau_{\text{p}}(\nu) &= L \left(\frac{\sqrt{1 - \left(\frac{\nu_p}{\nu}\right)^2}}{c} - \frac{1}{c} \right) \approx -\frac{L\:\nu_p^2}{2\:c} \cdot \frac{1}{\nu^2},
\end{align}
where $L = \SI{2.5}{\giga\m}$ denotes the LISA armlength. PRN and sideband signals propagate at the group velocity, hence they are delayed by the group delay:
\begin{align}
	\Delta \tau_{\text{g}}^{\text{prn}} &= \Delta \tau_{\text{g}}(\SI{281}{\tera\hertz}\pm\SI{1}{\mega\hertz}) \approx \SI{12.7}{\pico\m},\\
	\Delta \tau_{\text{g}}^{\text{sb}} &= \Delta \tau_{\text{g}}(\SI{281}{\tera\hertz}\pm\SI{2.4}{\giga \hertz}) \approx \SI{12.7}{\pico\m}.
\end{align}
The phase delay is negative, because the phase velocity is bigger than $c$. Therefore, the laser phase is advanced with respect to a wave propagating in vacuum. For the LISA carrier this phase advancement corresponds to
\begin{align}
	\Delta \tau_{\text{p}}(\SI{281}{\tera\hertz}) \approx -\SI{12.7}{\pico\m}.
\end{align}

\bibliography{references}

\begin{thebibliography}{32}%
\makeatletter
\providecommand \@ifxundefined [1]{%
 \@ifx{#1\undefined}
}%
\providecommand \@ifnum [1]{%
 \ifnum #1\expandafter \@firstoftwo
 \else \expandafter \@secondoftwo
 \fi
}%
\providecommand \@ifx [1]{%
 \ifx #1\expandafter \@firstoftwo
 \else \expandafter \@secondoftwo
 \fi
}%
\providecommand \natexlab [1]{#1}%
\providecommand \enquote  [1]{``#1''}%
\providecommand \bibnamefont  [1]{#1}%
\providecommand \bibfnamefont [1]{#1}%
\providecommand \citenamefont [1]{#1}%
\providecommand \href@noop [0]{\@secondoftwo}%
\providecommand \href [0]{\begingroup \@sanitize@url \@href}%
\providecommand \@href[1]{\@@startlink{#1}\@@href}%
\providecommand \@@href[1]{\endgroup#1\@@endlink}%
\providecommand \@sanitize@url [0]{\catcode `\\12\catcode `\$12\catcode
  `\&12\catcode `\#12\catcode `\^12\catcode `\_12\catcode `\%12\relax}%
\providecommand \@@startlink[1]{}%
\providecommand \@@endlink[0]{}%
\providecommand \url  [0]{\begingroup\@sanitize@url \@url }%
\providecommand \@url [1]{\endgroup\@href {#1}{\urlprefix }}%
\providecommand \urlprefix  [0]{URL }%
\providecommand \Eprint [0]{\href }%
\providecommand \doibase [0]{https://doi.org/}%
\providecommand \selectlanguage [0]{\@gobble}%
\providecommand \bibinfo  [0]{\@secondoftwo}%
\providecommand \bibfield  [0]{\@secondoftwo}%
\providecommand \translation [1]{[#1]}%
\providecommand \BibitemOpen [0]{}%
\providecommand \bibitemStop [0]{}%
\providecommand \bibitemNoStop [0]{.\EOS\space}%
\providecommand \EOS [0]{\spacefactor3000\relax}%
\providecommand \BibitemShut  [1]{\csname bibitem#1\endcsname}%
\let\auto@bib@innerbib\@empty
\bibitem [{\citenamefont {Amaro-Seoane}\ \emph {et~al.}(2017)\citenamefont
  {Amaro-Seoane}, \citenamefont {Audley}, \citenamefont {Babak}, \citenamefont
  {Baker}, \citenamefont {Barausse}, \citenamefont {Bender}, \citenamefont
  {Berti}, \citenamefont {Binetruy}, \citenamefont {Born}, \citenamefont
  {Bortoluzzi} \emph {et~al.}}]{Amaro2017LISA}%
  \BibitemOpen
  \bibfield  {author} {\bibinfo {author} {\bibfnamefont {P.}~\bibnamefont
  {Amaro-Seoane}}, \bibinfo {author} {\bibfnamefont {H.}~\bibnamefont
  {Audley}}, \bibinfo {author} {\bibfnamefont {S.}~\bibnamefont {Babak}},
  \bibinfo {author} {\bibfnamefont {J.}~\bibnamefont {Baker}}, \bibinfo
  {author} {\bibfnamefont {E.}~\bibnamefont {Barausse}}, \bibinfo {author}
  {\bibfnamefont {P.}~\bibnamefont {Bender}}, \bibinfo {author} {\bibfnamefont
  {E.}~\bibnamefont {Berti}}, \bibinfo {author} {\bibfnamefont
  {P.}~\bibnamefont {Binetruy}}, \bibinfo {author} {\bibfnamefont
  {M.}~\bibnamefont {Born}}, \bibinfo {author} {\bibfnamefont {D.}~\bibnamefont
  {Bortoluzzi}}, \emph {et~al.},\ }\bibfield  {title} {\bibinfo {title} {Laser
  interferometer space antenna},\ }\href@noop {} {\bibfield  {journal}
  {\bibinfo  {journal} {arXiv preprint arXiv:1702.00786}\ } (\bibinfo {year}
  {2017})}\BibitemShut {NoStop}%
\bibitem [{\citenamefont {Gerberding}\ \emph {et~al.}(2013)\citenamefont
  {Gerberding}, \citenamefont {Sheard}, \citenamefont {Bykov}, \citenamefont
  {Kullmann}, \citenamefont {Delgado}, \citenamefont {Danzmann},\ and\
  \citenamefont {Heinzel}}]{Gerberding:Phasemeter}%
  \BibitemOpen
  \bibfield  {author} {\bibinfo {author} {\bibfnamefont {O.}~\bibnamefont
  {Gerberding}}, \bibinfo {author} {\bibfnamefont {B.}~\bibnamefont {Sheard}},
  \bibinfo {author} {\bibfnamefont {I.}~\bibnamefont {Bykov}}, \bibinfo
  {author} {\bibfnamefont {J.}~\bibnamefont {Kullmann}}, \bibinfo {author}
  {\bibfnamefont {J.~J.~E.}\ \bibnamefont {Delgado}}, \bibinfo {author}
  {\bibfnamefont {K.}~\bibnamefont {Danzmann}},\ and\ \bibinfo {author}
  {\bibfnamefont {G.}~\bibnamefont {Heinzel}},\ }\bibfield  {title} {\bibinfo
  {title} {Phasemeter core for intersatellite laser heterodyne interferometry:
  modelling, simulations and experiments},\ }\href@noop {} {\bibfield
  {journal} {\bibinfo  {journal} {Classical and Quantum Gravity}\ }\textbf
  {\bibinfo {volume} {30}},\ \bibinfo {pages} {235029} (\bibinfo {year}
  {2013})}\BibitemShut {NoStop}%
\bibitem [{\citenamefont {Barke}\ \emph {et~al.}(2014)\citenamefont {Barke},
  \citenamefont {Brause}, \citenamefont {Bykov}, \citenamefont
  {Esteban~Delgado}, \citenamefont {Enggaard}, \citenamefont {Gerberding},
  \citenamefont {Heinzel}, \citenamefont {Kullmann}, \citenamefont {Pedersen},\
  and\ \citenamefont {Rasmussen}}]{barkeLISAMetrologySystem}%
  \BibitemOpen
  \bibfield  {author} {\bibinfo {author} {\bibfnamefont {S.}~\bibnamefont
  {Barke}}, \bibinfo {author} {\bibfnamefont {N.}~\bibnamefont {Brause}},
  \bibinfo {author} {\bibfnamefont {I.}~\bibnamefont {Bykov}}, \bibinfo
  {author} {\bibfnamefont {J.~J.}\ \bibnamefont {Esteban~Delgado}}, \bibinfo
  {author} {\bibfnamefont {A.}~\bibnamefont {Enggaard}}, \bibinfo {author}
  {\bibfnamefont {O.}~\bibnamefont {Gerberding}}, \bibinfo {author}
  {\bibfnamefont {G.}~\bibnamefont {Heinzel}}, \bibinfo {author} {\bibfnamefont
  {J.}~\bibnamefont {Kullmann}}, \bibinfo {author} {\bibfnamefont {S.~M.}\
  \bibnamefont {Pedersen}},\ and\ \bibinfo {author} {\bibfnamefont
  {T.}~\bibnamefont {Rasmussen}},\ }\bibfield  {title} {\bibinfo {title}
  {L{ISA} metrology system-final report},\ }\href@noop {} {\  (\bibinfo {year}
  {2014})}\BibitemShut {NoStop}%
\bibitem [{\citenamefont {Armstrong}\ \emph {et~al.}(1999)\citenamefont
  {Armstrong}, \citenamefont {Estabrook},\ and\ \citenamefont
  {Tinto}}]{ArmstrongTDI}%
  \BibitemOpen
  \bibfield  {author} {\bibinfo {author} {\bibfnamefont {J.}~\bibnamefont
  {Armstrong}}, \bibinfo {author} {\bibfnamefont {F.}~\bibnamefont
  {Estabrook}},\ and\ \bibinfo {author} {\bibfnamefont {M.}~\bibnamefont
  {Tinto}},\ }\bibfield  {title} {\bibinfo {title} {Time-delay interferometry
  for space-based gravitational wave searches},\ }\href@noop {} {\bibfield
  {journal} {\bibinfo  {journal} {The Astrophysical Journal}\ }\textbf
  {\bibinfo {volume} {527}},\ \bibinfo {pages} {814} (\bibinfo {year}
  {1999})}\BibitemShut {NoStop}%
\bibitem [{\citenamefont {Tinto}\ \emph {et~al.}(2002)\citenamefont {Tinto},
  \citenamefont {Estabrook},\ and\ \citenamefont
  {Armstrong}}]{TintoTDIforLISA}%
  \BibitemOpen
  \bibfield  {author} {\bibinfo {author} {\bibfnamefont {M.}~\bibnamefont
  {Tinto}}, \bibinfo {author} {\bibfnamefont {F.~B.}\ \bibnamefont
  {Estabrook}},\ and\ \bibinfo {author} {\bibfnamefont {J.}~\bibnamefont
  {Armstrong}},\ }\bibfield  {title} {\bibinfo {title} {Time-delay
  interferometry for {LISA}},\ }\href@noop {} {\bibfield  {journal} {\bibinfo
  {journal} {Physical Review D}\ }\textbf {\bibinfo {volume} {65}},\ \bibinfo
  {pages} {082003} (\bibinfo {year} {2002})}\BibitemShut {NoStop}%
\bibitem [{\citenamefont {Hartwig}\ \emph {et~al.}(2022)\citenamefont
  {Hartwig}, \citenamefont {Bayle}, \citenamefont {Staab}, \citenamefont
  {Hees}, \citenamefont {Lilley},\ and\ \citenamefont
  {Wolf}}]{Hartwig:TDIwoSync}%
  \BibitemOpen
  \bibfield  {author} {\bibinfo {author} {\bibfnamefont {O.}~\bibnamefont
  {Hartwig}}, \bibinfo {author} {\bibfnamefont {J.-B.}\ \bibnamefont {Bayle}},
  \bibinfo {author} {\bibfnamefont {M.}~\bibnamefont {Staab}}, \bibinfo
  {author} {\bibfnamefont {A.}~\bibnamefont {Hees}}, \bibinfo {author}
  {\bibfnamefont {M.}~\bibnamefont {Lilley}},\ and\ \bibinfo {author}
  {\bibfnamefont {P.}~\bibnamefont {Wolf}},\ }\bibfield  {title} {\bibinfo
  {title} {Time-delay interferometry without clock synchronization},\ }\href
  {https://doi.org/10.1103/PhysRevD.105.122008} {\bibfield  {journal} {\bibinfo
   {journal} {Phys. Rev. D}\ }\textbf {\bibinfo {volume} {105}},\ \bibinfo
  {pages} {122008} (\bibinfo {year} {2022})}\BibitemShut {NoStop}%
\bibitem [{\citenamefont {Esteban}\ \emph {et~al.}(2009)\citenamefont
  {Esteban}, \citenamefont {Bykov}, \citenamefont {Mar{\'\i}n}, \citenamefont
  {Heinzel},\ and\ \citenamefont {Danzmann}}]{EstebanPRN1}%
  \BibitemOpen
  \bibfield  {author} {\bibinfo {author} {\bibfnamefont {J.~J.}\ \bibnamefont
  {Esteban}}, \bibinfo {author} {\bibfnamefont {I.}~\bibnamefont {Bykov}},
  \bibinfo {author} {\bibfnamefont {A.~F.~G.}\ \bibnamefont {Mar{\'\i}n}},
  \bibinfo {author} {\bibfnamefont {G.}~\bibnamefont {Heinzel}},\ and\ \bibinfo
  {author} {\bibfnamefont {K.}~\bibnamefont {Danzmann}},\ }\bibfield  {title}
  {\bibinfo {title} {Optical ranging and data transfer development for
  {LISA}},\ }in\ \href@noop {} {\emph {\bibinfo {booktitle} {Journal of
  Physics: Conference Series}}},\ Vol.\ \bibinfo {volume} {154}\ (\bibinfo
  {organization} {IOP Publishing},\ \bibinfo {year} {2009})\ p.\ \bibinfo
  {pages} {012025}\BibitemShut {NoStop}%
\bibitem [{\citenamefont {Esteban}\ \emph {et~al.}(2010)\citenamefont
  {Esteban}, \citenamefont {Garc{\'\i}a}, \citenamefont {Eichholz},
  \citenamefont {Peinado}, \citenamefont {Bykov}, \citenamefont {Heinzel},\
  and\ \citenamefont {Danzmann}}]{EstebanPRN2}%
  \BibitemOpen
  \bibfield  {author} {\bibinfo {author} {\bibfnamefont {J.~J.}\ \bibnamefont
  {Esteban}}, \bibinfo {author} {\bibfnamefont {A.~F.}\ \bibnamefont
  {Garc{\'\i}a}}, \bibinfo {author} {\bibfnamefont {J.}~\bibnamefont
  {Eichholz}}, \bibinfo {author} {\bibfnamefont {A.~M.}\ \bibnamefont
  {Peinado}}, \bibinfo {author} {\bibfnamefont {I.}~\bibnamefont {Bykov}},
  \bibinfo {author} {\bibfnamefont {G.}~\bibnamefont {Heinzel}},\ and\ \bibinfo
  {author} {\bibfnamefont {K.}~\bibnamefont {Danzmann}},\ }\bibfield  {title}
  {\bibinfo {title} {Ranging and phase measurement for {LISA}},\ }in\
  \href@noop {} {\emph {\bibinfo {booktitle} {Journal of Physics: Conference
  Series}}},\ Vol.\ \bibinfo {volume} {228}\ (\bibinfo {organization} {IOP
  Publishing},\ \bibinfo {year} {2010})\ p.\ \bibinfo {pages}
  {012045}\BibitemShut {NoStop}%
\bibitem [{\citenamefont {Heinzel}\ \emph {et~al.}(2011)\citenamefont
  {Heinzel}, \citenamefont {Esteban}, \citenamefont {Barke}, \citenamefont
  {Otto}, \citenamefont {Wang}, \citenamefont {Garcia},\ and\ \citenamefont
  {Danzmann}}]{Heinzel-Ranging}%
  \BibitemOpen
  \bibfield  {author} {\bibinfo {author} {\bibfnamefont {G.}~\bibnamefont
  {Heinzel}}, \bibinfo {author} {\bibfnamefont {J.~J.}\ \bibnamefont
  {Esteban}}, \bibinfo {author} {\bibfnamefont {S.}~\bibnamefont {Barke}},
  \bibinfo {author} {\bibfnamefont {M.}~\bibnamefont {Otto}}, \bibinfo {author}
  {\bibfnamefont {Y.}~\bibnamefont {Wang}}, \bibinfo {author} {\bibfnamefont
  {A.~F.}\ \bibnamefont {Garcia}},\ and\ \bibinfo {author} {\bibfnamefont
  {K.}~\bibnamefont {Danzmann}},\ }\bibfield  {title} {\bibinfo {title}
  {Auxiliary functions of the {LISA} laser link: ranging, clock noise transfer
  and data communication},\ }\href@noop {} {\bibfield  {journal} {\bibinfo
  {journal} {Classical and Quantum Gravity}\ }\textbf {\bibinfo {volume}
  {28}},\ \bibinfo {pages} {094008} (\bibinfo {year} {2011})}\BibitemShut
  {NoStop}%
\bibitem [{\citenamefont {Hartwig}(2021)}]{hartwig2021instrumental}%
  \BibitemOpen
  \bibfield  {author} {\bibinfo {author} {\bibfnamefont {O.}~\bibnamefont
  {Hartwig}},\ }\bibfield  {title} {\bibinfo {title} {Instrumental modelling
  and noise reduction algorithms for the laser interferometer space antenna},\
  }\href@noop {} {\  (\bibinfo {year} {2021})}\BibitemShut {NoStop}%
\bibitem [{\citenamefont {Wang}\ \emph {et~al.}(2014)\citenamefont {Wang},
  \citenamefont {Heinzel},\ and\ \citenamefont {Danzmann}}]{Wang:KF}%
  \BibitemOpen
  \bibfield  {author} {\bibinfo {author} {\bibfnamefont {Y.}~\bibnamefont
  {Wang}}, \bibinfo {author} {\bibfnamefont {G.}~\bibnamefont {Heinzel}},\ and\
  \bibinfo {author} {\bibfnamefont {K.}~\bibnamefont {Danzmann}},\ }\bibfield
  {title} {\bibinfo {title} {First stage of {LISA} data processing: Clock
  synchronization and arm-length determination via a hybrid-extended {K}alman
  filter},\ }\href@noop {} {\bibfield  {journal} {\bibinfo  {journal} {Physical
  Review D}\ }\textbf {\bibinfo {volume} {90}},\ \bibinfo {pages} {064016}
  (\bibinfo {year} {2014})}\BibitemShut {NoStop}%
\bibitem [{\citenamefont {Sutton}\ \emph {et~al.}(2010)\citenamefont {Sutton},
  \citenamefont {McKenzie}, \citenamefont {Ware},\ and\ \citenamefont
  {Shaddock}}]{sutton2010laser}%
  \BibitemOpen
  \bibfield  {author} {\bibinfo {author} {\bibfnamefont {A.}~\bibnamefont
  {Sutton}}, \bibinfo {author} {\bibfnamefont {K.}~\bibnamefont {McKenzie}},
  \bibinfo {author} {\bibfnamefont {B.}~\bibnamefont {Ware}},\ and\ \bibinfo
  {author} {\bibfnamefont {D.~A.}\ \bibnamefont {Shaddock}},\ }\bibfield
  {title} {\bibinfo {title} {Laser ranging and communications for {LISA}},\
  }\href@noop {} {\bibfield  {journal} {\bibinfo  {journal} {Optics Express}\
  }\textbf {\bibinfo {volume} {18}},\ \bibinfo {pages} {20759} (\bibinfo {year}
  {2010})}\BibitemShut {NoStop}%
\bibitem [{\citenamefont {Tinto}\ \emph {et~al.}(2005)\citenamefont {Tinto},
  \citenamefont {Vallisneri},\ and\ \citenamefont {Armstrong}}]{TintoTDIR}%
  \BibitemOpen
  \bibfield  {author} {\bibinfo {author} {\bibfnamefont {M.}~\bibnamefont
  {Tinto}}, \bibinfo {author} {\bibfnamefont {M.}~\bibnamefont {Vallisneri}},\
  and\ \bibinfo {author} {\bibfnamefont {J.}~\bibnamefont {Armstrong}},\
  }\bibfield  {title} {\bibinfo {title} {Time-delay interferometric ranging for
  space-borne gravitational-wave detectors},\ }\href@noop {} {\bibfield
  {journal} {\bibinfo  {journal} {Physical Review D}\ }\textbf {\bibinfo
  {volume} {71}},\ \bibinfo {pages} {041101} (\bibinfo {year}
  {2005})}\BibitemShut {NoStop}%
\bibitem [{\citenamefont {Bayle}\ \emph {et~al.}(2021)\citenamefont {Bayle},
  \citenamefont {Hartwig},\ and\ \citenamefont
  {Staab}}]{Bayle:TDI-in-Frequency}%
  \BibitemOpen
  \bibfield  {author} {\bibinfo {author} {\bibfnamefont {J.-B.}\ \bibnamefont
  {Bayle}}, \bibinfo {author} {\bibfnamefont {O.}~\bibnamefont {Hartwig}},\
  and\ \bibinfo {author} {\bibfnamefont {M.}~\bibnamefont {Staab}},\ }\bibfield
   {title} {\bibinfo {title} {Adapting time-delay interferometry for {LISA}
  data in frequency},\ }\href@noop {} {\bibfield  {journal} {\bibinfo
  {journal} {Physical Review D}\ }\textbf {\bibinfo {volume} {104}},\ \bibinfo
  {pages} {023006} (\bibinfo {year} {2021})}\BibitemShut {NoStop}%
\bibitem [{\citenamefont {Bayle}\ and\ \citenamefont
  {Hartwig}(2023)}]{Bayle:LISA-Simulation}%
  \BibitemOpen
  \bibfield  {author} {\bibinfo {author} {\bibfnamefont {J.-B.}\ \bibnamefont
  {Bayle}}\ and\ \bibinfo {author} {\bibfnamefont {O.}~\bibnamefont
  {Hartwig}},\ }\bibfield  {title} {\bibinfo {title} {Unified model for the
  {LISA} measurements and instrument simulations},\ }\href@noop {} {\bibfield
  {journal} {\bibinfo  {journal} {Physical Review D}\ }\textbf {\bibinfo
  {volume} {107}},\ \bibinfo {pages} {083019} (\bibinfo {year}
  {2023})}\BibitemShut {NoStop}%
\bibitem [{\citenamefont {Brzozowski}\ \emph {et~al.}(2022)\citenamefont
  {Brzozowski}, \citenamefont {Robertson}, \citenamefont {Fitzsimons},
  \citenamefont {Ward}, \citenamefont {Keogh}, \citenamefont {Taylor},
  \citenamefont {Milanova}, \citenamefont {Perreur-Lloyd}, \citenamefont {Ali},
  \citenamefont {Earle} \emph {et~al.}}]{BrzozowskiLISA-OB}%
  \BibitemOpen
  \bibfield  {author} {\bibinfo {author} {\bibfnamefont {W.}~\bibnamefont
  {Brzozowski}}, \bibinfo {author} {\bibfnamefont {D.}~\bibnamefont
  {Robertson}}, \bibinfo {author} {\bibfnamefont {E.}~\bibnamefont
  {Fitzsimons}}, \bibinfo {author} {\bibfnamefont {H.}~\bibnamefont {Ward}},
  \bibinfo {author} {\bibfnamefont {J.}~\bibnamefont {Keogh}}, \bibinfo
  {author} {\bibfnamefont {A.}~\bibnamefont {Taylor}}, \bibinfo {author}
  {\bibfnamefont {M.}~\bibnamefont {Milanova}}, \bibinfo {author}
  {\bibfnamefont {M.}~\bibnamefont {Perreur-Lloyd}}, \bibinfo {author}
  {\bibfnamefont {Z.}~\bibnamefont {Ali}}, \bibinfo {author} {\bibfnamefont
  {A.}~\bibnamefont {Earle}}, \emph {et~al.},\ }\bibfield  {title} {\bibinfo
  {title} {The {LISA} optical bench: an overview and engineering challenges},\
  }\href@noop {} {\bibfield  {journal} {\bibinfo  {journal} {Space Telescopes
  and Instrumentation 2022: Optical, Infrared, and Millimeter Wave}\ }\textbf
  {\bibinfo {volume} {12180}},\ \bibinfo {pages} {211} (\bibinfo {year}
  {2022})}\BibitemShut {NoStop}%
\bibitem [{\citenamefont {Barranco}\ and\ \citenamefont
  {Heinzel}(2021)}]{BarrancoQPD}%
  \BibitemOpen
  \bibfield  {author} {\bibinfo {author} {\bibfnamefont {G.~F.}\ \bibnamefont
  {Barranco}}\ and\ \bibinfo {author} {\bibfnamefont {G.}~\bibnamefont
  {Heinzel}},\ }\bibfield  {title} {\bibinfo {title} {A {DC}-coupled,
  {HBT}-based transimpedance amplifier for the {LISA} quadrant
  photoreceivers},\ }\href {https://doi.org/10.1109/TAES.2021.3068437}
  {\bibfield  {journal} {\bibinfo  {journal} {IEEE Transactions on Aerospace
  and Electronic Systems}\ }\textbf {\bibinfo {volume} {57}},\ \bibinfo {pages}
  {2899} (\bibinfo {year} {2021})}\BibitemShut {NoStop}%
\bibitem [{\citenamefont {Euringer}\ \emph {et~al.}(2024)\citenamefont
  {Euringer}, \citenamefont {Hechenblaikner}, \citenamefont {Soualle},\ and\
  \citenamefont {Fichter}}]{Euringer-Code-Tracking}%
  \BibitemOpen
  \bibfield  {author} {\bibinfo {author} {\bibfnamefont {P.}~\bibnamefont
  {Euringer}}, \bibinfo {author} {\bibfnamefont {G.}~\bibnamefont
  {Hechenblaikner}}, \bibinfo {author} {\bibfnamefont {F.}~\bibnamefont
  {Soualle}},\ and\ \bibinfo {author} {\bibfnamefont {W.}~\bibnamefont
  {Fichter}},\ }\bibfield  {title} {\bibinfo {title} {Performance analysis of
  sequential carrier- and code-tracking receivers in the context of
  high-precision spaceborne metrology systems},\ }\href
  {https://doi.org/10.1109/TIM.2023.3332388} {\bibfield  {journal} {\bibinfo
  {journal} {IEEE Transactions on Instrumentation and Measurement}\ }\textbf
  {\bibinfo {volume} {73}},\ \bibinfo {pages} {1} (\bibinfo {year}
  {2024})}\BibitemShut {NoStop}%
\bibitem [{\citenamefont {Barke}(2015)}]{Barke2015Thesis}%
  \BibitemOpen
  \bibfield  {author} {\bibinfo {author} {\bibfnamefont {S.}~\bibnamefont
  {Barke}},\ }\href@noop {} {\emph {\bibinfo {title} {Inter-spacecraft
  frequency distribution for future gravitational wave observatories}}}\
  (\bibinfo  {publisher} {Hannover: Gottfried Wilhelm Leibniz Universit{\"a}t
  Hannover},\ \bibinfo {year} {2015})\BibitemShut {NoStop}%
\bibitem [{\citenamefont {Otto}(2015)}]{Otto2015TDI}%
  \BibitemOpen
  \bibfield  {author} {\bibinfo {author} {\bibfnamefont {M.}~\bibnamefont
  {Otto}},\ }\bibfield  {title} {\bibinfo {title} {Time-delay interferometry
  simulations for the laser interferometer space antenna},\ }\href@noop {} {\
  (\bibinfo {year} {2015})}\BibitemShut {NoStop}%
\bibitem [{\citenamefont {Martens}\ and\ \citenamefont
  {Joffre}(2021)}]{MartensOD}%
  \BibitemOpen
  \bibfield  {author} {\bibinfo {author} {\bibfnamefont {W.}~\bibnamefont
  {Martens}}\ and\ \bibinfo {author} {\bibfnamefont {E.}~\bibnamefont
  {Joffre}},\ }\bibfield  {title} {\bibinfo {title} {Trajectory design for the
  {ESA} {LISA} mission},\ }\href@noop {} {\bibfield  {journal} {\bibinfo
  {journal} {The Journal of the Astronautical Sciences}\ }\textbf {\bibinfo
  {volume} {68}},\ \bibinfo {pages} {402} (\bibinfo {year} {2021})}\BibitemShut
  {NoStop}%
\bibitem [{\citenamefont {Hees}\ \emph {et~al.}(2014)\citenamefont {Hees},
  \citenamefont {Bertone},\ and\ \citenamefont {Le~Poncin-Lafitte}}]{HeesLTT}%
  \BibitemOpen
  \bibfield  {author} {\bibinfo {author} {\bibfnamefont {A.}~\bibnamefont
  {Hees}}, \bibinfo {author} {\bibfnamefont {S.}~\bibnamefont {Bertone}},\ and\
  \bibinfo {author} {\bibfnamefont {C.}~\bibnamefont {Le~Poncin-Lafitte}},\
  }\bibfield  {title} {\bibinfo {title} {Relativistic formulation of coordinate
  light time, doppler, and astrometric observables up to the second
  post-{M}inkowskian order},\ }\href@noop {} {\bibfield  {journal} {\bibinfo
  {journal} {Physical Review D}\ }\textbf {\bibinfo {volume} {89}},\ \bibinfo
  {pages} {064045} (\bibinfo {year} {2014})}\BibitemShut {NoStop}%
\bibitem [{\citenamefont {Bayle}\ \emph
  {et~al.}(2022{\natexlab{a}})\citenamefont {Bayle}, \citenamefont {Hartwig},\
  and\ \citenamefont {Staab}}]{bayle_jean_baptiste_2022_7071251}%
  \BibitemOpen
  \bibfield  {author} {\bibinfo {author} {\bibfnamefont {J.-B.}\ \bibnamefont
  {Bayle}}, \bibinfo {author} {\bibfnamefont {O.}~\bibnamefont {Hartwig}},\
  and\ \bibinfo {author} {\bibfnamefont {M.}~\bibnamefont {Staab}},\ }\href
  {https://doi.org/10.5281/zenodo.7071251} {\bibinfo {title} {L{ISA}
  {I}nstrument}} (\bibinfo {year} {2022}{\natexlab{a}})\BibitemShut {NoStop}%
\bibitem [{\citenamefont {Bayle}\ \emph
  {et~al.}(2022{\natexlab{b}})\citenamefont {Bayle}, \citenamefont {Hartwig},
  \citenamefont {Petiteau},\ and\ \citenamefont
  {Lilley}}]{bayle_jean_baptiste_2022_6461078}%
  \BibitemOpen
  \bibfield  {author} {\bibinfo {author} {\bibfnamefont {J.-B.}\ \bibnamefont
  {Bayle}}, \bibinfo {author} {\bibfnamefont {O.}~\bibnamefont {Hartwig}},
  \bibinfo {author} {\bibfnamefont {A.}~\bibnamefont {Petiteau}},\ and\
  \bibinfo {author} {\bibfnamefont {M.}~\bibnamefont {Lilley}},\ }\href
  {https://doi.org/10.5281/zenodo.6461078} {\bibinfo {title} {L{ISAN}ode}}
  (\bibinfo {year} {2022}{\natexlab{b}})\BibitemShut {NoStop}%
\bibitem [{\citenamefont {Bayle}\ \emph
  {et~al.}(2022{\natexlab{c}})\citenamefont {Bayle}, \citenamefont {Hees},
  \citenamefont {Lilley}, \citenamefont {Le~Poncin-Lafitte}, \citenamefont
  {Martens},\ and\ \citenamefont {Joffre}}]{bayle_jean_baptiste_2022_7700361}%
  \BibitemOpen
  \bibfield  {author} {\bibinfo {author} {\bibfnamefont {J.-B.}\ \bibnamefont
  {Bayle}}, \bibinfo {author} {\bibfnamefont {A.}~\bibnamefont {Hees}},
  \bibinfo {author} {\bibfnamefont {M.}~\bibnamefont {Lilley}}, \bibinfo
  {author} {\bibfnamefont {C.}~\bibnamefont {Le~Poncin-Lafitte}}, \bibinfo
  {author} {\bibfnamefont {W.}~\bibnamefont {Martens}},\ and\ \bibinfo {author}
  {\bibfnamefont {E.}~\bibnamefont {Joffre}},\ }\href
  {https://doi.org/10.5281/zenodo.7700361} {\bibinfo {title} {L{ISA} {O}rbits}}
  (\bibinfo {year} {2022}{\natexlab{c}})\BibitemShut {NoStop}%
\bibitem [{\citenamefont {Staab}\ \emph {et~al.}(2023)\citenamefont {Staab},
  \citenamefont {Bayle},\ and\ \citenamefont
  {Hartwig}}]{staab_martin_2023_7704609}%
  \BibitemOpen
  \bibfield  {author} {\bibinfo {author} {\bibfnamefont {M.}~\bibnamefont
  {Staab}}, \bibinfo {author} {\bibfnamefont {J.-B.}\ \bibnamefont {Bayle}},\
  and\ \bibinfo {author} {\bibfnamefont {O.}~\bibnamefont {Hartwig}},\ }\href
  {https://doi.org/10.5281/zenodo.7704609} {\bibinfo {title} {Py{TDI}}}
  (\bibinfo {year} {2023})\BibitemShut {NoStop}%
\bibitem [{\citenamefont {Tinto}\ \emph {et~al.}(2003)\citenamefont {Tinto},
  \citenamefont {Shaddock}, \citenamefont {Sylvestre},\ and\ \citenamefont
  {Armstrong}}]{TintoTDIAccuracy}%
  \BibitemOpen
  \bibfield  {author} {\bibinfo {author} {\bibfnamefont {M.}~\bibnamefont
  {Tinto}}, \bibinfo {author} {\bibfnamefont {D.~A.}\ \bibnamefont {Shaddock}},
  \bibinfo {author} {\bibfnamefont {J.}~\bibnamefont {Sylvestre}},\ and\
  \bibinfo {author} {\bibfnamefont {J.}~\bibnamefont {Armstrong}},\ }\bibfield
  {title} {\bibinfo {title} {Implementation of time-delay interferometry for
  lisa},\ }\href@noop {} {\bibfield  {journal} {\bibinfo  {journal} {Physical
  Review D}\ }\textbf {\bibinfo {volume} {67}},\ \bibinfo {pages} {122003}
  (\bibinfo {year} {2003})}\BibitemShut {NoStop}%
\bibitem [{\citenamefont {Francis}\ \emph {et~al.}(2015)\citenamefont
  {Francis}, \citenamefont {Shaddock}, \citenamefont {Sutton}, \citenamefont
  {De~Vine}, \citenamefont {Ware}, \citenamefont {Spero}, \citenamefont
  {Klipstein},\ and\ \citenamefont {McKenzie}}]{francis2015tone}%
  \BibitemOpen
  \bibfield  {author} {\bibinfo {author} {\bibfnamefont {S.~P.}\ \bibnamefont
  {Francis}}, \bibinfo {author} {\bibfnamefont {D.~A.}\ \bibnamefont
  {Shaddock}}, \bibinfo {author} {\bibfnamefont {A.~J.}\ \bibnamefont
  {Sutton}}, \bibinfo {author} {\bibfnamefont {G.}~\bibnamefont {De~Vine}},
  \bibinfo {author} {\bibfnamefont {B.}~\bibnamefont {Ware}}, \bibinfo {author}
  {\bibfnamefont {R.~E.}\ \bibnamefont {Spero}}, \bibinfo {author}
  {\bibfnamefont {W.~M.}\ \bibnamefont {Klipstein}},\ and\ \bibinfo {author}
  {\bibfnamefont {K.}~\bibnamefont {McKenzie}},\ }\bibfield  {title} {\bibinfo
  {title} {Tone-assisted time delay interferometry on {GRACE} {F}ollow-{O}n},\
  }\href@noop {} {\bibfield  {journal} {\bibinfo  {journal} {Physical Review
  D}\ }\textbf {\bibinfo {volume} {92}},\ \bibinfo {pages} {012005} (\bibinfo
  {year} {2015})}\BibitemShut {NoStop}%
\bibitem [{\citenamefont {Baghi}\ \emph {et~al.}(2023)\citenamefont {Baghi},
  \citenamefont {Baker}, \citenamefont {Slutsky},\ and\ \citenamefont
  {Thorpe}}]{Baghi2023-PCA}%
  \BibitemOpen
  \bibfield  {author} {\bibinfo {author} {\bibfnamefont {Q.}~\bibnamefont
  {Baghi}}, \bibinfo {author} {\bibfnamefont {J.~G.}\ \bibnamefont {Baker}},
  \bibinfo {author} {\bibfnamefont {J.}~\bibnamefont {Slutsky}},\ and\ \bibinfo
  {author} {\bibfnamefont {J.~I.}\ \bibnamefont {Thorpe}},\ }\bibfield  {title}
  {\bibinfo {title} {Fully data-driven time-delay interferometry with
  time-varying delays},\ }\href@noop {} {\bibfield  {journal} {\bibinfo
  {journal} {Annalen der Physik}\ ,\ \bibinfo {pages} {2200447}} (\bibinfo
  {year} {2023})}\BibitemShut {NoStop}%
\bibitem [{\citenamefont {Yamamoto}\ \emph {et~al.}(2022)\citenamefont
  {Yamamoto}, \citenamefont {Vorndamme}, \citenamefont {Hartwig}, \citenamefont
  {Staab}, \citenamefont {Schwarze},\ and\ \citenamefont
  {Heinzel}}]{Yamamoto:HexagonSync}%
  \BibitemOpen
  \bibfield  {author} {\bibinfo {author} {\bibfnamefont {K.}~\bibnamefont
  {Yamamoto}}, \bibinfo {author} {\bibfnamefont {C.}~\bibnamefont {Vorndamme}},
  \bibinfo {author} {\bibfnamefont {O.}~\bibnamefont {Hartwig}}, \bibinfo
  {author} {\bibfnamefont {M.}~\bibnamefont {Staab}}, \bibinfo {author}
  {\bibfnamefont {T.~S.}\ \bibnamefont {Schwarze}},\ and\ \bibinfo {author}
  {\bibfnamefont {G.}~\bibnamefont {Heinzel}},\ }\bibfield  {title} {\bibinfo
  {title} {Experimental verification of intersatellite clock synchronization at
  {LISA} performance levels},\ }\href@noop {} {\bibfield  {journal} {\bibinfo
  {journal} {Physical Review D}\ }\textbf {\bibinfo {volume} {105}},\ \bibinfo
  {pages} {042009} (\bibinfo {year} {2022})}\BibitemShut {NoStop}%
\bibitem [{\citenamefont {Schwarze}\ \emph {et~al.}(2019)\citenamefont
  {Schwarze}, \citenamefont {Barranco}, \citenamefont {Penkert}, \citenamefont
  {Kaufer}, \citenamefont {Gerberding},\ and\ \citenamefont
  {Heinzel}}]{Schwarze:Hexagon}%
  \BibitemOpen
  \bibfield  {author} {\bibinfo {author} {\bibfnamefont {T.~S.}\ \bibnamefont
  {Schwarze}}, \bibinfo {author} {\bibfnamefont {G.~F.}\ \bibnamefont
  {Barranco}}, \bibinfo {author} {\bibfnamefont {D.}~\bibnamefont {Penkert}},
  \bibinfo {author} {\bibfnamefont {M.}~\bibnamefont {Kaufer}}, \bibinfo
  {author} {\bibfnamefont {O.}~\bibnamefont {Gerberding}},\ and\ \bibinfo
  {author} {\bibfnamefont {G.}~\bibnamefont {Heinzel}},\ }\bibfield  {title}
  {\bibinfo {title} {Picometer-stable hexagonal optical bench to verify {LISA}
  phase extraction linearity and precision},\ }\href@noop {} {\bibfield
  {journal} {\bibinfo  {journal} {Physical review letters}\ }\textbf {\bibinfo
  {volume} {122}},\ \bibinfo {pages} {081104} (\bibinfo {year}
  {2019})}\BibitemShut {NoStop}%
\bibitem [{\citenamefont {Smetana}(2020)}]{Smetana:Solar-Wind}%
  \BibitemOpen
  \bibfield  {author} {\bibinfo {author} {\bibfnamefont {A.}~\bibnamefont
  {Smetana}},\ }\bibfield  {title} {\bibinfo {title} {Background for
  gravitational wave signal at {LISA} from refractive index of solar wind
  plasma},\ }\href@noop {} {\bibfield  {journal} {\bibinfo  {journal} {Monthly
  Notices of the Royal Astronomical Society: Letters}\ }\textbf {\bibinfo
  {volume} {499}},\ \bibinfo {pages} {L77} (\bibinfo {year}
  {2020})}\BibitemShut {NoStop}%
\end{thebibliography}%
\end{document}